\documentclass[12pt]{article}
\pdfoutput=1
\usepackage{makeidx}
\makeindex
\usepackage[a4paper]{geometry}
\usepackage{jheppub,amsmath,amssymb,amsfonts,amsxtra,mathrsfs,graphics,graphicx,amsthm,epsfig,ytableau,bm,longtable,float,color,tikz,mathtools,xfrac,footnote,rotating,lscape}

\usepackage[all]{xy}
\usetikzlibrary{decorations.pathmorphing}
\usetikzlibrary{decorations.markings}
\usetikzlibrary{arrows, decorations.markings, calc, fadings, decorations.pathreplacing, patterns, decorations.pathmorphing, positioning}

\usetikzlibrary{positioning,shapes}
\usetikzlibrary{chains}
\usetikzlibrary{arrows,fit,decorations.pathreplacing}
\tikzstyle{every picture}+=[remember picture]
\tikzstyle{na} = [baseline=-.5ex]

\makeatletter
\newcommand{\vast}{\bBigg@{2}}
\newcommand{\Vast}{\bBigg@{3}}
\makeatother

\usepackage{calligra}
\DeclareMathAlphabet{\mathcalligra}{T1}{calligra}{m}{n}

\newcommand{\re}{\,\mathbb{R}\mbox{e}\,}

\newcommand{\ie}{\textit{i.e.}}

\numberwithin{equation}{section}

\newcommand{\nn}{\nonumber}

\newcommand{\be}{\begin{equation}} \newcommand{\ee}{\end{equation}}
\newcommand{\bea}{\begin{equation} \begin{aligned}} \newcommand{\eea}{\end{aligned} \end{equation}}

\def\tilde{\widetilde}

\def\hat{\widehat}

\def\bar{\overline}

\def\rt2{\sqrt{2}}

\def\Tr{\mathop{\rm Tr}}

\def\CG{{\cal G}}
\def\CH{{\cal H}}

\def\CN{{\cal N}}

\def\CV{{\cal V}}

\def\1{{\ds 1}}

\newcommand{\cC}{\mathcal{C}}

\newcommand{\cN}{\mathcal{N}}

\newcommand{\cV}{\mathcal{V}}

\newcommand{\bC}{\mathbb{C}}

\newcommand{\fg}{\mathfrak{g}}
\newcommand{\fm}{\mathfrak{m}}
\newcommand{\fh}{\mathfrak{h}}
\newcommand{\fn}{\mathfrak{n}}

\newcommand{\Li}{{\rm Li}}

\def\SO{\mathrm{SO}}

\def\SU{\mathrm{SU}}
\def\SL{\mathrm{SL}}

\def\repa{\raise4pt\hbox{$\square$}\mkern-14mu\raise-4pt\hbox{$\square$}}
\def\repab{\overline{\raise4pt\hbox{$\square$}\mkern-14mu\raise-4pt\hbox{$\square$}\mkern-1mu}}

\def\smileface{\ensuremath{\hbox{\large$\bigcirc$}\mkern-15mu\raise-1pt\hbox{\scriptsize$\smallsmile$}%
\mkern-10mu\raise4pt\hbox{..}\mkern4mu}}
\def\frownface{\ensuremath{\hbox{\large$\bigcirc$}\mkern-15mu\raise-1pt\hbox{\scriptsize$\smallfrown$}%
\mkern-10mu\raise4pt\hbox{..}\mkern4mu}}

\def\node#1#2{\overset{#1}{\underset{#2}{\circ}}}

\def\SO{\mathrm{SO}}

\def\U{\mathrm{U}}
\def\SU{\mathrm{SU}}
\def\SL{\mathrm{SL}}

\newcommand{\ds}{\displaystyle}
\newcommand{\wb}{\overline}

\newcommand{\fR}{\mathfrak{R}}

\newcommand{\ft}{\mathfrak{t}}

\newcommand{\ba}{\begin{array}}
\newcommand{\ea}{\end{array}}
\newcommand{\bi}{\begin{itemize}}
\newcommand{\ei}{\end{itemize}}

\def\bea#1\eea{\allowdisplaybreaks \begin{align}#1\end{align}}
 \newcommand{\ben}{\begin{enumerate}}
\newcommand{\een}{\end{enumerate}}
\newcommand{\bean}{\begin{eqnarray*}}
\newcommand{\eean}{\end{eqnarray*}}
\newcommand{\eref}[1]{(\ref{#1})}

\newcommand{\tQ}{\widetilde{Q}}

\newcommand{\BC}{\mathbb{C}}
\newcommand{\BR}{\mathbb{R}}

\newcommand{\BZ}{\mathbb{Z}}

\newcommand{\comment}[1]{}

\newcommand{\Sym}{\mathrm{Sym}}

\title{Large $N$ topologically twisted index: necklace quivers, dualities, and Sasaki-Einstein spaces}

\author[a,b]{Seyed Morteza Hosseini}
\author[c]{and Noppadol Mekareeya}

\affiliation[a]{Dipartimento di Fisica, Universit\`a di Milano-Bicocca, \\I-20126 Milano, Italy}
\affiliation[b]{INFN, sezione di Milano-Bicocca, I-20126 Milano, Italy}
\affiliation[c]{Theory Department, CERN, \\CH-1211, Geneva 23, Switzerland}

\emailAdd{morteza.hosseini@mib.infn.it}
\emailAdd{noppadol.mekareeya@cern.ch}

\preprint{CERN-TH-2016-083}

\abstract{In this paper, we calculate the topological free energy for a number of $\cN \geq 2$ Yang-Mills-Chern-Simons-matter theories at large $N$ and fixed Chern-Simons levels.
The topological free energy is defined as the logarithm of the partition function of the theory on $S^2 \times S^1$ with a topological A-twist along $S^2$ and can be reduced to a matrix integral by exploiting the localization technique.
The theories of our interest are dual to a variety of Calabi-Yau four-fold singularities, including a product of two asymptotically locally Euclidean singularities and the cone over various well-known homogeneous Sasaki-Einstein seven-manifolds, $N^{0,1,0}$, $V^{5,2}$, and $Q^{1,1,1}$.
We check that the large $N$ topological free energy can be matched for theories which are related by dualities, including mirror symmetry and $\mathrm{SL}(2,\mathbb{Z})$ duality.
}

\begin{document}

\setcounter{tocdepth}{3}
\maketitle

%*******************************************************************************
%
%     Main body
%
%*******************************************************************************

\date{Dated: \today}

\section{Introduction}

For three-dimensional field theories with $\cN = 2$ supersymmetry, the partition function of theories on $S_{\rm A}^2 \times S^1$, with a topological A-twist along $S^2$ \cite{Witten:1991zz},
is reduced to a matrix integral which depends on background magnetic fluxes $\fn_I$ and fugacities (chemical potentials) $y_I\, (\Delta_I)$ for the flavor symmetries of the theory \cite{Benini:2015noa}.
It is explicitly given by a contour integral of a meromorphic form, where the position of the poles of the meromorphic integrand is governed by a set of algebraic equations, called the Bethe ansatz equations (BAEs) \cite{Benini:2015eyy}. The latter can also be found by extremizing a ``Bethe potential" functional.
Upon dimensional reduction on $S^2$, the matrix model can be interpreted as the Witten index
\be
Z(\fn_I, \Delta_I) = \Tr (-1)^F e^{-\beta H} e^{i J_I \Delta_I} \, ,
\ee
of the $\cN = 2$ supersymmetric quantum mechanics, where $J_I$ are the generators of the flavor symmetries.

A recent evaluation of the twisted matrix model for the $\cN = 6$ $\U(N)_k \times \U(N)_{-k}$ ABJM theory at large $N$ and fixed Chern-Simons levels $k$, describing $N$ M2-branes on $\BC^4/\BZ_k$ \cite{Aharony:2008ug}, showed that the index scales as $N^{3/2}$ and it reads \cite{Benini:2015eyy}
\bea \label{freeABJM}
\mathfrak{F}_{\text{ABJM}_k} = -\frac{ k^{1/2} N^{3/2}}{3} \sqrt{ 2 \prod_{i=1}^{2} \Delta_{A_i} \Delta_{B_i}}
\sum_{i=1}^{2}\left(\frac{\fn_{A_i}}{\Delta_{A_i}} + \frac{\fn_{B_i}}{\Delta_{B_i}}\right)\, .
\eea
Here, $\mathfrak{F}_{\text{ABJM}_k}$ is the topological free energy $\mathfrak{F} = \re \log Z$ of the ABJM theory. We have also denoted the chemical potentials of the bi-fundamental fields $A_{i}$, $B_{i}$ transforming in the $({\bf N},\overline{\bf N})$ and $(\overline{\bf N},{\bf N})$ of the two gauge groups, by $\Delta_{A_i}$, $\Delta_{B_i}$ and their corresponding flavor magnetic fluxes by $\fn_{A_{i}}$, $\fn_{B_i}$.
The topological free energy precisely reproduces the entropy of the magnetically charged BPS black holes in ${\rm AdS}_4 \times S^7$ \cite{Benini:2015eyy}.

The topologically twisted index is a powerful tool to investigate the properties of three-dimensional $\cN \geq 2$ gauge theories \cite{Benini:2015noa, Benini:2015eyy}.
In this paper, we present the large $N$ limit of the topologically twisted index for a number of Yang-Mills-Chern-Simons-matter quiver theories with $\CN\geq 2$ supersymmetry.
We provide explicit solutions to the BAEs at large $N$ and compute the topological free energy.
In particular, we match the topological free energy between theories which are related to each other by dualities, including mirror symmetry \cite{Intriligator:1996ex} and $\SL(2,\BZ)$ duality \cite{Aharony:1997ju, Gaiotto:2008ak, Assel:2014awa}.
Moreover, we consider quiver gauge theories which are thought to describe the low energy dynamics of a stack of M2-branes probing a CY$_{4}$ singularity.

We start by studying quiver gauge theories that can be realized on M2-branes probing two asymptotically locally Euclidean (ALE) singularities \cite{Porrati:1996xi}.
These include the ADHM \cite{Atiyah:1978ri} and the Kronheimer-Nakajima \cite{kronheimer1990yang} quivers, as well as some of the necklace quiver theories considered in \cite{Imamura:2008nn}.
We show that the topological free energy of such theories can be written as that of the ABJM theory times a numerical factor,
which depends on the orders of the ALE singularities and the Chern-Simons level of the ABJM theory.

We then switch to the analysis of theories proposed as dual to the M-theory backgrounds ${\rm AdS}_4 \times Y_7$, where $Y_7$ is a homogeneous Sasaki-Einstein manifold.
In particular, we compute the topological free energy for $N^{0,1,0}$ with $\cN = 3$ and $V^{5,2}$, $Q^{1,1,1}$ with $\cN = 2$ supersymmetry \cite{Gaiotto:2009tk, Imamura:2011uj, Cheon:2011th, Martelli:2009ga, Jafferis:2009th, Benini:2009qs,Cremonesi:2010ae}. One of the features of these geometries compared to ${\rm AdS}_4 \times S^7$ background is the existence of nontrivial two-cycles in the Sasaki-Einstein manifold, which are identified with the baryons in the dual quiver gauge theory \cite{Gubser:1998bc,Witten:1998xy}.

The plan of this paper is as follows.
In Section \ref{sec:prelim} we review the topologically twisted index.
In particular, the rules for constructing the Bethe potential and the twisted matrix model at large $N$, which are derived in \cite{Hosseini:2016tor} are summarized in this section.

In Section \ref{sec:N4susy} we discuss quiver gauge theories with $\CN=4$ supersymmetry.
The solution to the BAEs  of these theories are particularly simple and hence serve as pedagogical examples before moving on to more complicated models.

In Section \ref{sec:N3susy} we focus on $\CN=3$ necklace quiver theories that can be obtained from certain $\CN=4$ theories by turning on Chern-Simons couplings to some of the gauge groups \cite{Aharony:1997ju, Gaiotto:2008ak, Imamura:2008nn, Assel:2012cj, Assel:2014awa}.
We also verify the matching of the topological free energy for theories which are $\SL(2, \BZ)$ dual to each other.
This section is ended with the discussion of the theory proposed to describe M2-branes on $N^{0,1,0}/\BZ_k$ \cite{Hohenegger:2009as,Hikida:2009tp,Gaiotto:2009tk}.

In Section \ref{sec:N2susy} we consider quiver Chern-Simons-matter theories with $\cN = 2$ supersymmetry.
The two models for $V^{5,2}$ proposed by \cite{Martelli:2009ga} and \cite{Jafferis:2009th} are discussed in this section and their topological free energy are matched.
We then proceed to theories which come from flavoring the $\cN = 6$ ABJM theory
and flavored variations of the three-dimensional $\cN = 8$ Yang-Mills theory \cite{Gaiotto:2009tk, Benini:2009qs,Cremonesi:2010ae}.
The conclusions and discussion are presented in Section \ref{sec:conclude}.

\section{The  topologically twisted index}  \label{sec:prelim}
We are interested in Yang-Mills-Chern-Simons quiver theories with (anti-)fundamental, adjoint, and \emph{non-chiral} bi-fundamental\footnote{For any bi-fundamental field transforming in the $({\bf N}, \overline{\bf N})$ representations of $\U(N)_a \times \U(N)_b$
there exists another bi-fundamental field transforming in the conjugate representation $(\overline{\bf N}, {\bf N})$.}
 matter fields and some number $|G|$ of $\U(N)^{(a)}$ gauge groups.
Let us introduce the holomorphic Cartan combinations $u = A_t + i \beta \sigma$ on the complexified Cartan subalgebra $\fg_\bC$,
where $A_t$ is a Wilson line on $S^1$ and runs over the maximal torus of the gauge group $G$, $\sigma$ is the real scalar in the vector multiplet and runs over the corresponding Cartan subalgebra,
and $\beta$ is the radius of $S^1$. We denote the Chern-Simons coupling by $k$.
Given a weight $\rho_I$ of the representation $\fR_I$, we use a notation where $x^{\rho_I} = e^{ i \rho_{I}(u) }$.
The localized twisted index on the $S_{\rm A}^2 \times S^1$ background reads \cite{Benini:2015noa},
\be
\label{path}
Z (\fn, y) = \frac1{|\mathfrak{W}|} \; \sum_{\fm \,\in\, \Gamma_\fh} \; \oint_\cC \;   \prod_{\text{Cartan}} \left (\frac{dx}{2\pi i x}  x^{k \fm} \right ) \prod_{\alpha \in G} (1-x^\alpha) \,  \prod_I \prod_{\rho_I \in \fR_I} \bigg( \frac{x^{\rho_I/2} \, y_I^{1/2}}{1-x^{\rho_I} \, y_I} \bigg)^{\rho_I(\fm)- \fn_I  +1}  \, .
\ee
Here, $\alpha$ are the roots of $G$ and $|\mathfrak{W}|$ denotes the order of the Weyl group.

In this paper, we consider theories of which the R-symmetry can mix with any other abelian global symmetry such that the R-charges can be chosen to be integral-valued. The fugacities $y_I$ and flavor magnetic fluxes $\fn_I$ are parameterized by the global symmetries of the theory, subject to the conditions discussed in the next section.

The index is integrated over the zero-mode gauge variables $x=e^{i u}$ and summed over gauge magnetic fluxes $\fm$, living in the co-root lattice $\Gamma_\fh$ of $G$.
A  $\U(1)$ topological symmetry with fugacity $\xi =e^{i \Delta_m}$ and magnetic flux $\ft$ contributes to the index as
\be
\label{topological}
Z^\text{top} = x^\ft \, \xi^\fm \, .
\ee

\subsection{Review of the large $N$ limit}

In this section we briefly review the large $N$ limit of the topologically twisted index which is derived in \cite{Hosseini:2016tor}.
Generalizing the results of \cite{Benini:2015eyy}, we consider the following large $N$ expansion for the eigenvalue distribution,
\be
u_i^{(a)} = i N^{1/2} t_i + v_i^{(a)} (t) + \ldots \, .
\ee
In the large $N$ limit, we define a density
\be
\rho(t) = \frac{1}{N} \sum_{i=1}^{N} \delta (t - t_i) \, ,
\ee
which becomes an integrable function in the continuum limit, satisfying
\be
\int dt\, \rho(t) = 1\, , \qquad \rho(t) \geq 0 \quad \text{pointwise.}
\ee
The position of the poles of the meromorphic integrand \eqref{path} is then found by extremizing a {\it Bethe potential} $\cV[\rho(t), v_a(t)]$.
We will impose the normalization of the density by introducing a Lagrange multiplier $\mu$.

We work in the M-theory limit where $N$ is large at fixed Chern-Simons level $k_a$.
We require the Chern-Simons levels sum to zero, \ie\,$\sum_{a=1}^{|G|} k_a = 0$, and hence the supergravity scaling $N^{3/2}$ is recovered.
Moreover, we only consider quiver gauge theories with \emph{non-chiral} bi-fundamental matter fields.
We also demand that the total number of fundamental fields equals the total number of anti-fundamental fields in the theory.
As it was shown in \cite{Hosseini:2016tor}, there are long-range forces come from the interactions between the eigenvalues.
In general, the long-range forces on $u_i^{(a)}$ in the Bethe potential cancel out only when
\be\label{no long-range forces Bethe}
\sum_{I \in a} \left(\pi - \Delta_I \right) \in 2 \pi \BZ \, ,
\ee
where the sum is taken over all bi-fundamental fields with one leg in the node $a$.\footnote{One should count adjoint fields twice.}
To have long-range forces cancellation in the index we should impose the following constraint
\be\label{no long-range forces free energy}
\sum_{I \in a} \left(1 - \fn_I \right) = 2 \, .
\ee

For quiver gauge theories which meet the above conditions, the large $N$ Bethe potential can be written as
\begin{align} \label{genbethepot}
& \frac{\cV \left[ \rho(t), v_a(t) \right]}{i N^{3/2}} = - \int dt\, t\, \rho(t)\, \sum_{a=1}^{|G|} \left( k_a v_a(t) + \Delta_m^{(a)} \right) \nn \\
& + \frac{1}{2} \int dt\, |t|\, \rho(t) \left\{ \sum_{\substack{\text{anti-fund} \\ a }} \Big[v_a(t) - \big( \tilde \Delta_a - \pi \big)\Big]
   - \sum_{\substack{\text{fund} \\ a }} \Big[v_a(t) + \big( \Delta_a - \pi \big)\Big]\right\} \nn \\
& + \int dt\, \rho(t)^2 \sum_{\substack{\text{bi-funds} \\ (b,a) \text{ and } (a,b) }} \left[g_+\left(\delta v_{ba}(t) + \Delta_{(b,a)}\right) - g_-\left(\delta v_{ba}(t) - \Delta_{(a,b)}\right)\right] \, \nn \\
& - \frac{i}{N^{1/2}} \int dt\, \rho(t) \sum_{\substack{\text{bi-funds} \\ (b,a) \text{ and } (a,b)}}\bigg[  \Li_2 \left(e^{i \big(\delta v_{ba}(t)+ {\Delta}_{(b,a)}\big)}\right) - \Li_2 \left(e^{i \big(\delta v_{ba}(t)- {\Delta}_{(a,b)}\big)}\right)  \bigg]~,
\end{align}
where  $\Delta_m^{(a)}$ is the chemical potential associated with the topological symmetry of the $a$-th gauge group, as described around \eref{topological}.  The Bethe potential $\cV \left[ \rho(t), v_a(t) \right]$ has to be extremized as a functional of  $\rho(t)$ and $v_a(t)$'s under the constraint that $\rho(t)$ is a density.\footnote{In our notations, $\Delta_{(a,b)}$ is the chemical potential associated to the bi-fundamental field transforming in the $({\bf N}, \overline{\bf N})$ representation of $\U(N)_a \times \U(N)_b$. The contribution of an adjoint field is also obtained by setting $a=b$ in the sum over bi-fundamental fields and dividing it by an explicit factor of two.}
Here, 
\bea
\delta v_{ba} (t) \equiv v_b (t) - v_a (t) \, ,
\eea  
and, for the sake of brevity, we shall abbreviate $ \delta v(t) := \delta v_{ba} (t)$ in the following discussion.  We also introduced the polynomial functions
\begin{equation}\label{gp gm}
g_\pm(u) = \frac{u^3}6 \mp \frac\pi2 u^2 + \frac{\pi^2}3 u \;,\qquad\qquad g_\pm'(u) = \frac{u^2}2 \mp \pi u + \frac{\pi^2}3 \, .
\end{equation}
This formula was derived assuming the bi-fundamental fields fulfill
\be
\label{inequalities for delta v0}
0 < \delta v + \Delta_{(b,a)} < 2\pi \;,\qquad\qquad\qquad -2\pi < \delta v - \Delta_{(a,b)} < 0 \, .
\ee
Moreover, we assume that $0 < \Delta < 2 \pi$.

When $\delta v + \Delta_{(b,a)} = 0 \text{ or } 2 \pi$ $(\delta v - \Delta_{(b,a)} = - 2 \pi \text{ or } 0)$, it is crucial to take into account the last line of \eref{genbethepot}; see also the discussion around (2.68) of \cite{Benini:2015eyy}.
%the discrete nature of the eigenvalues shows up, and the equation of motion for $\delta v$ does not need to hold.
This gives contribution to the {\it tails regions} where $\delta v$ has exponentially small correction to the large $N$ constant value:
\be
\delta v (t) = - \Delta_{(b,a)} + e^{-N^{1/2} Y_{(b,a)}}\, , \qquad \delta v (t) = \Delta_{(a,b)} - e^{-N^{1/2} Y_{(a,b)}}\, , \qquad (\text{mod } 2\pi)\, .
\ee
An explicit example will be discussed in Section \ref{sec:necklacealtCS}.

The invariance of the superpotential under global symmetries, imposes the following constraints
\be \label{constraints}
\sum_{I \in W} \Delta_I \in 2 \pi \BZ \, , \qquad\qquad \sum_{I\in W} \fn_I = 2 \, , \quad \text{with}~ \fn_I \in \BZ~, 
\ee
where the sum is taken over all the fields in each monomial term $W$ in the superpotential.
As we will see in the upcoming sections, we can always find a solution to the BAEs for
\be \label{marginality}
\sum_{I \in W} \Delta_I = 2 \pi \, .
\ee
We call this the ``marginality condition'' of the superpotential. Moreover, in all theories discussed in this paper (except the $V^{5,2}/\BZ_k$ theory discussed in Section \ref{sec:V52}), we can find an {\it integral} solution to the second equality of \eref{constraints}; this ensures that there always exists a choice of the $R$-charges that take integral values.\footnote{We thank the referee of JHEP for emphasising this point to us.}
Nevertheless, for the $V^{5,2}/\BZ_k$, the quantisation condition $\fn_I \in \BZ$ can be easily satisfied by considering the theory on a higher genus Riemann surface $\Sigma_g$ times a circle \cite{Benini:2016hjo}.
We discuss this in detail in Section \ref{sec:V52}. 

There is also a solution for
\be
\sum_{I \in W} \Delta_I = \left( |W| - 1 \right) 2 \pi \, ,
\ee
where $|W|$ is the number of fields in each term of the superpotential.  However, using the discrete symmetry $y_I \to 1/y_I\, (\Delta_I \to 2 \pi - \Delta_I)$ of the index, it can be mapped to \eqref{marginality}.

Once we find a solution to the BAEs, we plug it back into
\begin{align} \label{genentropy}
& \frac{\mathfrak{F}}{N^{3/2}} = - \frac{|G| \pi^2}{3} \int dt\, \rho(t)^2 - \sum_{a=1}^{|G|} \ft_a \int dt\, t\, \rho(t)
+ \frac{1}{2} \int dt\, |t|\, \rho(t) \left[\sum_{\substack{\text{anti-funds} \\ a}} (\tilde\fn_a - 1) + \sum_{\substack{\text{funds} \\ a}} (\fn_a - 1)\right] \nn \\
& - \int dt\, \rho(t)^2 \sum_{\substack{\text{bi-funds} \\ (b,a) \text{ and } (a,b) }} \left[(\fn_{(b,a)}-1)\, g'_+ \left(\delta v(t) + \Delta_{(b,a)}\right) + (\fn_{(a,b)}-1)\, g'_- \left(\delta v(t) - \Delta_{(a,b)}\right)\right] \nn \\
& - \sum_{\substack{\text{bi-fund} \\ (b,a) }} \fn_{(b,a)} \int_{\delta v \approx - \Delta_{(b,a)} (\text{mod } 2\pi)}  dt \, \rho(t) Y_{(b,a)}
- \sum_{\substack{\text{bi-fund} \\ (a,b) }} \fn_{(a,b)} \int_{\delta v \approx  \Delta_{(a,b)} (\text{mod } 2\pi)}  dt \, \rho(t) Y_{(a,b)} \, ,
\end{align}
to compute the topological free energy, at large $N$, of three-dimensional $\cN \geq 2$ Yang-Mills-Chern-Simons-matter theories placed on $S_{\rm A}^2 \times S^1$.

It is also possible to calculate $\mathfrak{F}$ using the powerful {\it index theorem} of \cite{Hosseini:2016tor}.
The topological free energy of any $\cN \geq 2$ quiver Chern-Simons-matter-gauge theory which fulfills the conditions \eqref{no long-range forces Bethe}, \eqref{no long-range forces free energy}, and \eqref{marginality}, can be written as
\begin{equation}\label{index theorem}
 \mathfrak{F} = - \frac{2}{\pi} \, \wb{\mathcal{V}}(\Delta_I) \,
 - \sum_{I}\, \left[ \left(\fn_I - \frac{\Delta_I}{\pi}\right) \frac{\partial \wb{\mathcal{V}}(\Delta_I)}{\partial \Delta_I}
  \right] \, .
\end{equation}
Here, $\wb{\cV}$ is the extremal value of the Bethe potential functional \eqref{genbethepot},
\begin{equation}\label{virial theorem}
 \wb{\mathcal{V}}(\Delta_I) \equiv - i \mathcal{V} \Big|_{\text{BAEs}} = \frac{2}{3} \mu N^{3/2} \, ,
\end{equation}
where the second equality can be understood as a virial theorem for matrix models (see Appendix B of \cite{Gulotta:2011si}).

In the following sections we will calculate the topological free energy $\mathfrak{F}$ by evaluating the functional \eqref{genentropy} on the solution to the BAEs, and thus the index theorem serves as an independent check of our results.

\section{Quivers with $\CN=4$ supersymmetry} \label{sec:N4susy}
In this section, we consider two quiver gauge theories with $\CN=4$ supersymmetry. As pointed out in \cite{Porrati:1996xi}, each of these theories can be realized in the worldvolume of M2-branes probing $\BC^2/\BZ_{n_1} \times \BC^2/\BZ_{n_2}$, for some positive integers $n_1$ and $n_2$.  We show below that the topological free energy of such theories can be written as $\sqrt{n_1 n_2/k}$ times that of the ABJM theory with Chern-Simons levels $(+k,-k)$.
We also match the index of a pair of theories which are mirror dual \cite{Intriligator:1996ex} to each other.
This serves as a check of the validity of our results.

\subsection{The ADHM quiver}
We consider $\U(N)$ gauge theory with one adjoint and $r$ fundamental hypermultiplets, whose $\CN=4$ quiver is given by
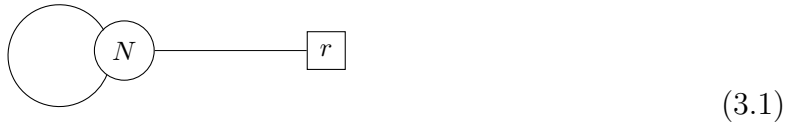
\bea \label{ADHMN4quiv}
\begin{tikzpicture}[font=\footnotesize, scale=0.9]
\begin{scope}[auto,%
  every node/.style={draw, minimum size=0.5cm}, node distance=2cm];
\node[circle]  (UN)  at (0.3,1.7) {$N$};
\node[rectangle, right=of UN] (Ur) {$r$};
\end{scope}
\draw(-0,2) arc (30:338:0.75cm);
\draw[black,solid,line width=0.1mm]  (UN) to (Ur) ; 
\end{tikzpicture}
\eea
where the circular node denotes the $\U(N)$ gauge group; the square node denotes the $\SU(r)$ flavor symmetry; the loop around the circular node denotes the adjoint hypermultiplet; and the line between $N$ and $r$ denotes the fundamental hypermultiplet.  The vacuum equations of the Higgs branch of the theory were used in the construction of the instanton solutions by Atiyah, Drinfeld, Hitchin and Manin \cite{Atiyah:1978ri}.  This quiver gauge theory hence acquires the name ``ADHM quiver''.

In $\CN=2$ notation, this theory contains three adjoint chiral fields: $\phi_{1}, \, \phi_2, \, \phi_3$, where $\phi_{1,2}$ come from the $\CN=4$ adjoint hypermultiplet and $\phi_3$ comes from the $\CN=4$ vector multiplet, and fundamental chiral fields $Q^i_a$, $\tQ^a_i$ with $a = 1, \ldots, N$ and $i=1, \ldots, r$.  The superpotential is
\bea
W= \tQ^i_a (\phi_3)^a_{~b} Q^b_i + (\phi_3)^a_{~b}[ \phi_1, \phi_2]^b_{~a} \, .
\eea
The $\CN=2$ quiver diagram is depicted below.
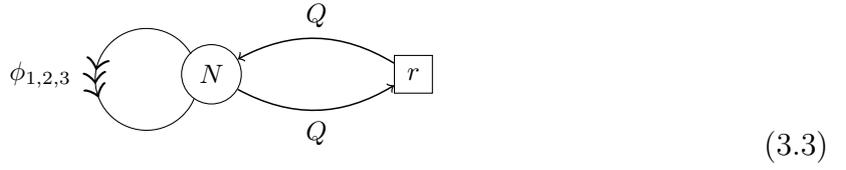
\bea
\begin{tikzpicture}[font=\footnotesize, scale=0.9]
\begin{scope}[auto,%
  every node/.style={draw, minimum size=0.5cm}, node distance=2cm];
\node[circle]  (UN)  at (0.3,1.7) {$N$};
\node[rectangle, right=of UN] (Ur) {$r$};
\end{scope}
\draw[decoration={markings, mark=at position 0.45 with {\arrow[scale=2.5]{>}}, mark=at position 0.5 with {\arrow[scale=2.5]{>}}, mark=at position 0.55 with {\arrow[scale=2.5]{>}}}, postaction={decorate}, shorten >=0.7pt] (-0,2) arc (30:340:0.75cm);
\draw[draw=black,solid,line width=0.2mm,->]  (UN) to[bend right=30] node[midway,below] {$Q$}node[midway,above] {}  (Ur) ; 
\draw[draw=black,solid,line width=0.2mm,<-]  (UN) to[bend left=30] node[midway,above] {$\tQ$} node[midway,above] {} (Ur) ;    
\node at (-2.2,1.7) {$\phi_{1,2,3}$};
\end{tikzpicture}
\eea
The Higgs branch of this gauge theory describes the moduli space of $N$ $\SU(r)$ instantons on $\BC^2$ \cite{Atiyah:1978ri} and the Coulomb branch is isomorphic to the space $\Sym^N (\BC^2/\BZ_r)$ \cite{deBoer:1996mp}.  This theory can be realized on the worldvolume of $N$ M2-branes probing $\BC^2 \times \BC^2/\BZ_r$ singularity \cite{Porrati:1996xi}.

\subsubsection{A solution to the system of BAEs} \label{sec:solnADHM2pi}
Let us denote, respectively, by $\Delta$, $\tilde{\Delta}$, $\Delta_{\phi_{1,2,3}}$ the chemical potentials associated to the flavor symmetries of $Q$, $\tQ$, $\phi_{1,2,3}$, and by $\fn$, $\tilde{\fn}$, ${\fn}_{\phi_{1,2,3}}$ the corresponding fluxes associated with their flavor symmetries.  We denote also by  $\Delta_m$ the chemical potential associated with the topological charge of the gauge group $\U(N)$.

The Bethe potential $\CV$ for this model can be obtained from \eref{genbethepot} as
\bea\label{ADHM Bethe}
\frac{\CV}{i N^{3/2}} & = \left( \sum_{i=1}^3 g_+(\Delta_{\phi_i})  \right) \int dt \, \rho(t)^2
-\frac{r}{2} \left[(\Delta - \pi) + (\tilde{\Delta} - \pi) \right] \int dt \, |t| \, \rho(t) \nn \\
& + \Delta_m \int dt \, t\, \rho(t) -\mu \left(\int dt \, \rho(t)-1 \right) \, .
\eea
Taking the variational derivative of $\CV$ with respect to $\rho(t)$, we obtain the BAE
\bea \label{BAEADHM1}
0 = 2 \rho(t) \sum_{i=1}^3 g_+(\Delta_{\phi_i}) - \frac{r}{2} |t| \left[ (\Delta - \pi) + (\tilde{\Delta} - \pi) \right] + \Delta_m t - \mu \, .
\eea

We first look for the solution satisfying the marginality condition on the superpotential, \ie,
\bea
{\Delta}+\tilde{\Delta} +{\Delta}_{\phi_3} = 2\pi\, , \qquad {\Delta}_{\phi_1}+{\Delta}_{\phi_2} +{\Delta}_{\phi_3} = 2\pi \, ,
\eea
and
\be
{\fn}+\tilde{\fn} +{\fn}_{\phi_3} = 2 \, , \qquad {\fn}_{\phi_1}+{\fn}_{\phi_2} +{\fn}_{\phi_3} = 2 \, . 
\ee

For later convenience, let us normalize the chemical potential associated with the topological charge as follows:
\bea
\chi = \frac{2}{r} \Delta_m \, . \label{defchiADHM}
\eea
Solving \eref{BAEADHM1}, we get
\bea
\rho(t) = \frac{2 \mu - r \Delta_{\phi_3}  \left| t\right| - r \chi t  }{2 \prod_{i=1}^3 \Delta_{\phi_i}} \, .
\eea
The solution is supported on the interval $[t_-,t_+]$ with $t_- < 0 < t_+$, where $t_\pm$ can be determined from $\rho(t_\pm)=0$:
\bea
 t_{\pm}= \pm \frac{2\mu }{(\Delta_{\phi_3} \pm \chi) r} \, .
\eea
The normalization $\int_{t_-}^{t_+} dt \, \rho(t) =1$ fixes
\be \label{ADHMmu}
\mu = \sqrt{\frac{r}{2} \Delta_{\phi_1} \Delta_{\phi_2}(\Delta_{\phi_3}+\chi)(\Delta_{\phi_3}-\chi)} \, .
\ee

\paragraph{The solution in the other ranges.}
Let us consider
\bea
{\Delta}+\tilde{\Delta} +{\Delta}_{\phi_3} = 2  \pi \ell \, , \qquad {\Delta}_{\phi_1}+{\Delta}_{\phi_2} +{\Delta}_{\phi_3} = 2 \pi \ell \, , \qquad \text{where $\ell \in \BZ_{\geq 0}$} \, .
\eea
For $\ell = 0$ and $\ell=3$, we have $\Delta=\tilde{\Delta}={\Delta}_{\phi_{1,2,3}}=0$ or $\Delta=\tilde{\Delta}={\Delta}_{\phi_{1,2,3}}=2\pi$, respectively. These are singular solutions.
For $\ell =2$, the solution can be mapped to the previous one (\ie{} $\ell = 1$) by a discrete symmetry
\bea
\Delta_I \rightarrow 2\pi - \Delta_I \,, ~\quad
\mu \rightarrow  - \mu\, , ~\quad \Delta_m \rightarrow - \Delta_m \, ,
\eea
where the index $I$ labels matter fields in the theory.
From now on, we shall consider only the solution satisfying the marginality condition \eqref{marginality}.

\subsubsection{The index at large $N$}
The topological free energy of the ADHM quiver can be derived from \eref{genentropy} as
\bea \label{index ADHM}
\frac{\mathfrak{F}_{\text{ADHM}}}{N^{3/2}} &= - \left[ \frac{\pi^2}{3} + \sum_{i=1}^3 (\fn_{\phi_i} -1) g'_+(\Delta_{\phi_i})  \right] \int dt \, \rho(t)^2 - \frac{r}{2} \ft \int dt \, t \, \rho(t) \nn \\
& + \frac{r}{2} \left[ \left(\fn -1 \right) + \left(\tilde{\fn} -1 \right) \right]  \int dt \, |t|\, \rho(t) \, ,
\eea
where $\ft$ is the magnetic flux conjugate to the variable $\chi$ defined in \eref{defchiADHM}.
Plugging the above solution back into \eqref{index ADHM}, we find that
\bea  \label{freeADHM}
\mathfrak{F}_{\text{ADHM}} = \sqrt{\frac{r}{k}}\, \mathfrak{F}_{\text{ABJM}_k} \, .
\eea
The map of the parameters is as follows,
\bea \label{paramADHM}
\Delta_{A_1} & = \frac{1}{2} (\Delta_{\phi_3}-\chi) \, , \qquad \Delta_{A_2} =  \frac{1}{2}(\Delta_{\phi_3}+\chi) \, , \qquad \Delta_{B_1}= \Delta_{\phi_1} \, , \qquad \Delta_{B_2} =\Delta_{\phi_2} \, , \nn \\
\fn_{A_1} & = \frac{1}{2} (\fn_{\phi_3}-\ft)\, , \qquad \fn_{A_2} =  \frac{1}{2}(\fn_{\phi_3}+\ft) \, , \qquad \fn_{B_1}= \fn_{\phi_1} \, , \qquad \fn_{B_2} =\fn_{\phi_2} \, .
\eea

The factor $\sqrt{r/k}$ in \eref{freeADHM} is the ratio between the orbifold order of $\Sym^N(\BC^2 \times \BC^2/\BZ_r)$ and that of $\Sym^N(\BC^2/\BZ_k)$; the former is the geometric branch of the ADHM theory and the latter is that of the ABJM theory with Chern-Simons levels $(+k, -k)$.

\subsection{The $A_{n-1}$ Kronheimer-Nakajima quiver}
We consider a necklace quiver with $\U(N)^n$ gauge group with a bi-fundamental hypermultiplet between the adjacent gauge groups and with $r$ flavors of fundamental hypermultiplets under the $n$-th gauge group.  The $\CN=4$ quiver is depicted below.
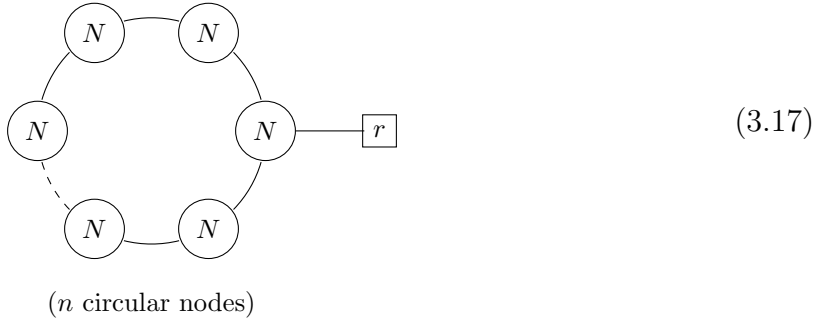
\bea \label{KNN4quiv}
\begin{tikzpicture}[baseline]
\def \n {6}
\def \radius {1.5cm}
\def \margin {16} % margin in angles, depends on the radius
\foreach \s in {1,...,5}
{
  \node[draw, circle] at ({360/\n * (\s - 2)}:\radius) {{\footnotesize $N$}};
  \draw[-, >=latex] ({360/\n * (\s - 3)+\margin}:\radius) 
    arc ({360/\n * (\s - 3)+\margin}:{360/\n * (\s-2)-\margin}:\radius);
}
\node[draw, circle] at ({360/3 * (3 - 1)}:\radius) {{\footnotesize $N$}};
\draw[dashed, >=latex] ({360/6 * (5 -2)+\margin}:\radius) 
    arc ({360/6 * (5 -2)+\margin}:{360/6 * (5-1)-\margin}:\radius);
\node[draw, rectangle] at (3,0) {{\footnotesize $r$}};
\draw[-, >=latex] (1+0.9,0) to (3-0.2,0);
\node[draw=none] at (0,-2.3) {{\footnotesize ($n$ circular nodes)}};
\end{tikzpicture}
\eea
As proposed by Kronheimer and Nakajima \cite{kronheimer1990yang}, the vacuum equations for the Higgs branch of this theory describes the hyperK\"ahler quotient of the moduli space of $\SU(r)$ instantons on $\BC^2/\BZ_n$ with $\SU(r)$ left unbroken by the monodromy at infinity. We shall henceforth refer to this quiver as the ``Kronheimer-Nakajima quiver''.

The corresponding $\CN=2$ quiver diagram is
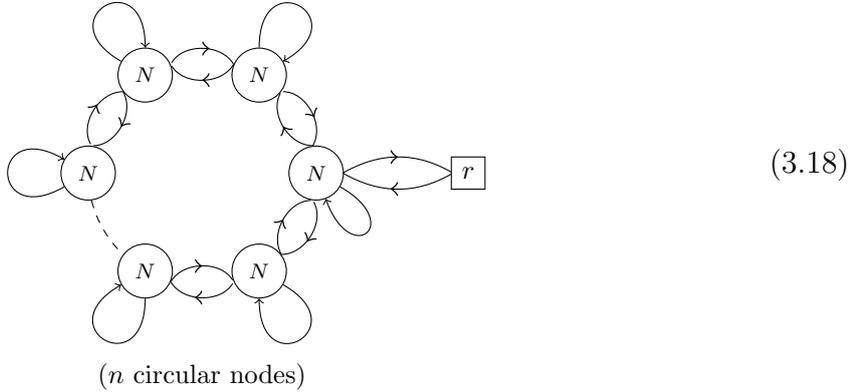
\bea \label{KNN2quiv}
\begin{tikzpicture}[baseline, font=\scriptsize]
\def \n {6}
\def \radius {1.5cm}
\def \margin {14} % margin in angles, depends on the radius
\foreach \s in {1,...,6}
{
%    \node[draw=none,minimum size=0.0001cm] at ({360/\n * (\s)}:\radius)   (\s) {$\tiny{\bullet}$};
  \node[draw, circle] at ({360/\n * (\s)}:\radius)  (\s) {$N$};
}
\foreach \s in {1,...,5}
{    
    \draw[black,-> ] (\s) edge [out={30+\s*60},in={-30+\s*60},loop,looseness=8] (\s);
} 
\draw[black,-> ] (6) edge [out={30+6*60-60},in={-30+6*60-40},loop,looseness=10] (6);
\foreach \s in {1,...,5}
{  
  \draw[decoration={markings, mark=at position 0.5 with {\arrow[scale=1.5]{>}}}, postaction={decorate}, shorten >=0.7pt]  ({360/\n * (\s - 3)+\margin}:\radius) to[bend left=60] node[midway,above] {}node[midway,below] {}  ({360/\n * (\s - 2)-\margin}:\radius) ; 
    \draw[decoration={markings, mark=at position 0.5 with {\arrow[scale=1.5]{<}}}, postaction={decorate}, shorten >=0.7pt]  ({360/\n * (\s - 3)+\margin}:\radius) to[bend right=60] node[midway,above] {}node[midway,below] {}  ({360/\n * (\s - 2)-\margin}:\radius) ;
}   
%\node[draw, circle] at ({360/3 * (3 - 1)}:\radius) {{\footnotesize $N$}};
\draw[dashed, >=latex] ({360/6 * (5 -2)+\margin}:\radius) 
    arc ({360/6 * (5 -2)+\margin}:{360/6 * (5-1)-\margin}:\radius);
\node[draw, rectangle] at (3.5,0) {{\footnotesize $r$}};
\draw[decoration={markings, mark=at position 0.5 with {\arrow[scale=1.5]{>}}}, postaction={decorate}, shorten >=0.7pt] (1.5+0.35,0)  to[bend left=30] node[midway,above] {}node[midway,below] {}   (3.5-0.2,0);
\draw[decoration={markings, mark=at position 0.5 with {\arrow[scale=1.5]{<}}}, postaction={decorate}, shorten >=0.7pt] (1.5+0.35,0)  to[bend right=30] node[midway,above] {}node[midway,below] {}   (3.5-0.2,0);
\node[draw=none] at (0,-2.7) {{\footnotesize ($n$ circular nodes)}};
\end{tikzpicture}
\eea
Let $Q_\alpha$ (with $\alpha=1,\ldots, n$) be the bi-fundamental field that goes from node $\alpha$ to node $\alpha+1$; $\tilde{Q}_\alpha$ be the bi-fundamental field that goes from node $\alpha+1$ to node $\alpha$; and $\phi_\alpha$ be the adjoint field under node $\alpha$.   Let us also denote by $q^i_{a}$ and $\tilde{q}^a_{i}$ the fundamental and anti-fundamental chiral multiplets under the $n$-th gauge group (with $a=1,\ldots, N$ and $i=1, \ldots, r$).  The superpotential is
\bea
W= \sum_{\alpha=1}^n  \Tr\left(Q_\alpha \phi_{\alpha+1} \tilde{Q}_\alpha- \tilde{Q}_{\alpha} \phi_{\alpha} Q_{\alpha}\right)+ \tilde{q}^a_i \, (\phi_n)^b_{~a}  \, {q}^i_b ~,
\eea
where we identify $\phi_{n+1}=\phi_1$. From now on, the index $\alpha$ labeling the nodes is taken modulo $n$ for any necklace quiver with $n$ nodes.

The Higgs branch of this gauge theory describes the moduli space of $N$ $\SU(r)$ instantons on $\BC^2/\BZ_n$ such that the monodromy at infinity preserves $\SU(r)$ symmetry \cite{kronheimer1990yang}, and the Coulomb branch describes the moduli space of $N$ $\SU(n)$ instantons on $\BC^2/\BZ_r$ such that the monodromy at infinity preserves $\SU(n)$ symmetry \cite{deBoer:1996mp, Porrati:1996xi, Witten:2009xu, Mekareeya:2015bla}.  It can be indeed realized on the worldvolume of $N$ M2-branes probing $\BC^2/\BZ_n \times \BC^2/\BZ_r$ singularity \cite{Porrati:1996xi}.  Note also that $3d$ mirror symmetry exchanges the Kronheimer-Nakajima quiver \eref{KNN4quiv} with $r=1$ and $n=2$ and the ADHM quiver \eref{ADHMN4quiv} with $r=2$. 

\subsubsection{A solution to the system of BAEs} \label{sec:solnKN}
Let us denote respectively by $\Delta_{Q_\alpha}$, $\Delta_{\tQ_\alpha}$, $\Delta_{\phi_{\alpha}}$, $\Delta_{q}$, $\Delta_{\tilde{q}}$ the chemical potentials associated to the flavor symmetries of $Q_{\alpha}$, $\tQ_{\alpha}$, $\phi_{\alpha}$, $q$ and $\tilde{q}$, and by $\fn_{Q_\alpha}$, ${\fn}_{\tQ_\alpha}$, ${\fn}_{\phi_{\alpha}}$, $\fn_q$, $\fn_{\tilde{q}}$ the corresponding fluxes associated with their flavor symmetries.  We also denote by $\Delta_m^{(\alpha)}$ the chemical potential associated with the topological charge for gauge group $\alpha$ and by $\ft^{(\alpha)}$ the associated magnetic flux.

From \eref{genbethepot}, the Bethe potential $\CV$ for this model is given by
\bea
\frac{\CV}{i N^{3/2}} & = \int dt \, \rho(t)^2 \sum_{\alpha=1}^n \left[ g_+ (\delta v^\alpha(t)+ {\Delta}_{\tQ_\alpha}) - g_- (\delta v^\alpha(t) - \Delta_{Q_\alpha})  + g_+(\Delta_{\phi_\alpha}) \right] \nn \\
& - \frac{r}{2} \left[ \left(\Delta_q - \pi \right) + \left({\Delta}_{\tilde{q}} - \pi \right) \right] \int dt \, |t| \, \rho(t)
+ \left( \sum_{\alpha =1}^n \Delta_m^{(\alpha)}  \right) \int dt \, t \, \rho(t) \nn \\
& -\mu \left(\int dt \, \rho(t)-1 \right) \, .
\eea
where $\delta v^{\alpha} = v^{\alpha+1} - v^\alpha$ and we identify $\delta v^{n+1} = \delta v^{1}$. Taking the variational derivatives of $\CV$ with respect to $\rho(t)$ and $\delta v^\alpha(t)$, we obtain the BAEs
\bea \label{BAEKN}
0 &= 2 \rho(t) \sum_{\alpha=1}^n \left[ g_+ (\delta v^\alpha(t)+ {\Delta}_{\tQ_\alpha}) - g_- (\delta v^\alpha(t) - \Delta_{Q_\alpha}) + g_+(\Delta_{\phi_\alpha})  \right]  \nn \\
& - \frac{r}{2} |t| \left[ (\Delta_q - \pi) + ({\Delta}_{\tilde{q}} - \pi) \right]
+ \left( \sum_{\alpha =1}^n \Delta_m^{(\alpha)}  \right) t - \mu \, ,\\
0&= \rho(t) \Big[ g'_+( \delta v^\alpha(t)+ \Delta_{\tQ \alpha})
- g'_- ( \delta v^\alpha(t) - \Delta_{Q \alpha}) \nn \\
& + g'_- ( \delta v^{\alpha-1}(t) - \Delta_{Q_{\alpha-1}})
- g'_+(\delta v^{\alpha-1}(t) + \Delta_{\tQ_{\alpha-1}}) \Big] \, , \qquad \alpha =1, \ldots, n \, .
\eea

The superpotential imposes the following constraints on the chemical potentials of the various fields:
\bea
{\Delta}_q+{\Delta}_{\tilde{q}} +{\Delta}_{\phi_n} = 2\pi \, , \quad {\Delta}_{Q_\alpha}+\Delta_{\phi_{\alpha+1}}+{\Delta}_{\tQ_\alpha} = 2\pi \, , \quad {\Delta}_{\tQ_\alpha}+\Delta_{\phi_{\alpha}}+{\Delta}_{Q_\alpha} = 2\pi \, . 
\eea

For notational convenience, we define
\bea\label{paraKN}
F_1= \sum_\alpha {\Delta}_{\tQ_\alpha}\, , \qquad F_3 = \Delta_{\phi_{n}} \, , \qquad {\Delta_m} = \frac{2}{r} \sum_{\alpha} \Delta_m^{(\alpha)}\, ,
\eea
and
\bea
F_2 = 2\pi -F_1-F_3 \, .
\eea
Solving the system of BAEs \eref{BAEKN}, we find that 
\bea
\rho(t) = \frac{2 \mu - r F_3  \left| t\right| -  r \Delta_m t  }{2 \prod_{i=1}^3 F_i} ~, \qquad \delta v^\alpha =  \frac{1}{n} F_1-\Delta_{\tQ_\alpha}~.
\eea
The support $[t_-, t_+]$ of $\rho(t)$ is determined by $\rho(t_\pm)=0$. We get
\bea
 t_\pm = \pm \frac{2\mu }{(F_3 \pm \Delta_m)r}\, .
\eea
The normalization $\int_{t_-}^{t_+} dt \, \rho(t) =1$ fixes
\be
\mu = \sqrt{\frac{3 r}{2} F_1 F_2 (F_3+\Delta_m) (F_3-\Delta_m)} \, . \label{muKN}
\ee

\subsubsection{The index at large $N$}

From \eref{genentropy}, the topological free energy of this quiver is given by
\bea \label{SKN}
\frac{\mathfrak{F}_{\text{KN}}}{N^{3/2}} & = - \frac{n \pi^2}{3} \int dt \, \rho(t)^2 - \left(\sum_{\alpha=1}^n \ft^{(\alpha)} \right)  \int dt \, t \, \rho(t)
+ \frac{r}{2} \left[ (\fn_q -1) + (\fn_{\tilde{q}} -1) \right]  \int dt \, |t|\, \rho(t) \nn \\
& -  \int dt \, \rho(t)^2\, \sum_{\alpha=1}^n \bigg[(\fn_{\tQ_\alpha} -1) g'_+(\delta v^\alpha(t)+ \Delta_{\tQ_\alpha})
+ (\fn_{Q_\alpha} -1) g'_-(\delta v^\alpha(t) - \Delta_{Q_\alpha}) \bigg] \nn \\
& - \sum_{\alpha=1}^n (\fn_{\phi_\alpha} -1) g'_+(\Delta_{\phi_\alpha}) \int dt \, \rho(t)^2
\eea
Plugging the above solution back into \eref{SKN}, we find that the topological free energy depends only on the parameters $F_1$, $F_2$, $F_3$ given by \eref{paraKN} and their corresponding conjugate charges
\bea
\fn_1= \sum_\alpha {\fn}_{\tQ_\alpha} \, , \qquad \fn_3 =\fn_{\phi_{n}} \, , \qquad {\ft} = \frac{2}{r} \sum_{\alpha} \ft^{(\alpha)}  \, .
\eea
Explicitly, we obtain
\bea \label{freeKN}
\mathfrak{F}_{\text{KN}} = \sqrt{\frac{nr}{k}}\, \mathfrak{F}_{\text{ABJM}_k}\, ,
\eea
with the following map of the parameters
\bea
\Delta_{A_1} &= \frac{1}{2} (F_3-\Delta_m) \, , \qquad \Delta_{A_2} =  \frac{1}{2}(F_3+\Delta_m) \, , \qquad \Delta_{B_1}= F_1 \, , \qquad \Delta_{B_2} =F_2 \, , \nn \\
\fn_{A_1} & = \frac{1}{2} (\fn_{3}-\ft) \, , \qquad \fn_{A_2} =  \frac{1}{2}(\fn_3+\ft) \, , \qquad \fn_{B_1}= \fn_{1} \, , \qquad \fn_{B_2} =\fn_{2} \, .
\eea
Notice that, this is completely analogous to that of the ADHM quiver presented in \eref{paramADHM}.

The factor $\sqrt{nr/k}$ in \eref{freeADHM} is the ratio between the product of the orbifold orders in $\Sym^N(\BC^2/\BZ_n \times \BC^2/\BZ_r)$ and that of $\Sym^N(\BC^2/\BZ_k)$, where the former is the geometric branch of the Kronheimer-Nakajima theory and the latter is that of the ABJM theory with Chern-Simons levels $(+k, -k)$.

\paragraph{Mirror symmetry \cite{Intriligator:1996ex}.}  The Kronheimer-Nakajima quiver \eref{KNN4quiv} with $r=1$ and $n=2$ is mirror dual to the ADHM quiver \eref{ADHMN4quiv} with $r=2$.  From \eref{freeADHM} and \eref{freeKN}, the topological free energy of the two theories are indeed equal:
\bea
\mathfrak{F}_{\text{KN}} \Big |_{r=1, \, n=2} = \mathfrak{F}_{\text{ADHM}} \Big |_{r=2} \, .
\eea

\section{Quivers with $\CN=3$ supersymmetry} \label{sec:N3susy}

A crucial difference between the theories considered in this section and those with $\CN=4$ supersymmetry is that the solution to the BAEs of the former are divided into several regions and the final result of the topological free energy comes from the sum of the contributions of each region.
Such a feature of the solution was already present in the ABJM theory and was discussed extensively in \cite{Benini:2015eyy}. In subsection \ref{altCSN3}, we deal with the necklace quiver with alternating Chern-Simons levels and present the Bethe potential, the BAEs and the procedure to solve them in detail.
The solutions for the other models in the following subsections can be derived in a similar fashion.

In subsections \ref{sec:necklacealtCS} and \ref{sec:necklacetwoCS}, we focus on theories whose geometric branch is a symmetric power of a product of two ALE singularities \cite{Imamura:2008nn, Cremonesi:2016nbo}.
Similarly to the preceding section, the topological free energy of such theories can be written as a numerical factor times the topological free energy of the ABJM theory, where the numerical factor equals to the square root of the ratio between the product of the orders of such singularities and the level of the ABJM theory.
Moreover, in a certain special case where the quiver is $\SL(2, \BZ)$ dual to a quiver with $\CN=4$ supersymmetry \cite{Gaiotto:2008ak, Assel:2012cj, Assel:2014awa, Cremonesi:2016nbo}, we match the topological free energy of two theories.

\subsection{The affine $A_{2m-1}$ quiver with alternating CS levels} \label{sec:necklacealtCS}
We are interested in the necklace quiver with $n=2m$ nodes, each with $\U(N)$ gauge group, and alternating Chern-Simons levels:
\bea \label{quivaltCS}
k_\alpha= \begin{cases} +k & \text{if $\alpha$ is odd} \\
-k & \text{if $\alpha$ is even} \end{cases}
\eea
The $\CN=2$ quiver diagram is depicted below.
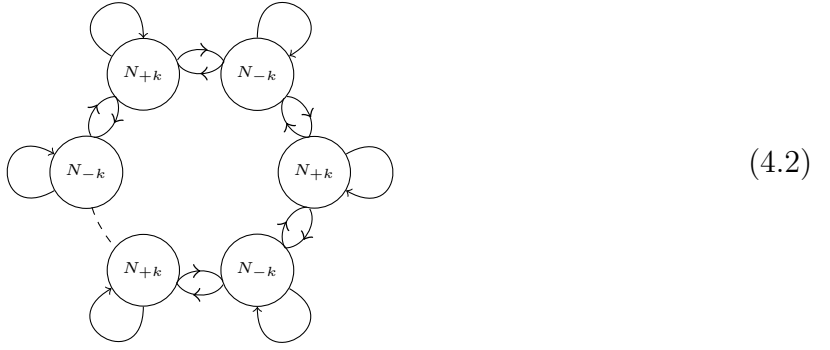
\bea
\begin{tikzpicture}[baseline,font=\tiny, every loop/.style={min distance=10mm,looseness=10}, every node/.style={minimum size=0.9cm}]
\def \n {6}
\def \radius {1.5cm}
\def \margin {18} % margin in angles, depends on the radius
\foreach \s in {1}
{
    \node[draw, circle] at ({2*360/\n * (\s)}:\radius)  (\s) {$N_{+k}$};
    \draw[black,-> ] (\s) edge [out={30+2*\s*60},in={-30+2*\s*60},loop,looseness=5] (\s);
} 
\foreach \s in {1}
{    
    \node[draw, circle]  at ({2*360/\n*\s-360/\n}:\radius) (\s)  {$N_{-k}$};
    \draw[black,-> ] (\s) edge [out={-30+2*\s*60},in={-90+2*\s*60},loop,looseness=5] (\s);
} 
\foreach \s in {2,3}
{
    \node[draw, circle] at ({2*360/\n * (\s)}:\radius)  (\s) {$N_{+k}$};
    \draw[black,-> ] (\s) edge [out={30+2*\s*60},in={-30+2*\s*60},loop,looseness=5] (\s);
} 
\foreach \s in {2,3}
{    
    \node[draw, circle]  at ({2*360/\n*\s-360/\n}:\radius) (\s)  {$N_{-k}$};
    \draw[black,-> ] (\s) edge [out={-30+2*\s*60},in={-90+2*\s*60},loop,looseness=5] (\s);
} 
\foreach \s in {1,...,5}
{  
  \draw[decoration={markings, mark=at position 0.5 with {\arrow[scale=1.5]{>}}}, postaction={decorate}, shorten >=0.7pt]  ({360/\n * (\s - 3)+\margin}:\radius) to[bend left=60] node[midway,above] {}node[midway,below] {}  ({360/\n * (\s - 2)-\margin}:\radius) ; 
    \draw[decoration={markings, mark=at position 0.5 with {\arrow[scale=1.5]{<}}}, postaction={decorate}, shorten >=0.7pt]  ({360/\n * (\s - 3)+\margin}:\radius) to[bend right=60] node[midway,above] {}node[midway,below] {}  ({360/\n * (\s - 2)-\margin}:\radius) ;
}   
\draw[dashed, >=latex] ({360/6 * (5 -2)+\margin}:\radius) 
    arc ({360/6 * (5 -2)+\margin}:{360/6 * (5-1)-\margin}:\radius);
%\node[draw=none] at (5,-1.5) {{\footnotesize ($\bullet$ = $U(N)$ gauge node; $n$ gauge nodes)}};
\end{tikzpicture}
\eea
Let $Q_\alpha$ be the bi-fundamental field that goes from node $\alpha$ to node $\alpha+1$; $\tilde{Q}_\alpha$ be the bi-fundamental field that goes from node $\alpha+1$ to node $\alpha$; and $\phi_\alpha$ be the adjoint field under node $\alpha$.  The superpotential can be written as
\bea
W=\sum_{\alpha=1}^n  \Tr\left(Q_\alpha \phi_{\alpha+1} \tilde{Q}_\alpha- \tilde{Q}_{\alpha} \phi_{\alpha} Q_{\alpha}\right) +\frac{k}{2} \sum_{\alpha=1}^m \Tr \left(\phi_{2\alpha-1}^2- \phi_{2\alpha}^2\right) \, .
\eea
After integrating out the massive adjoint fields, we have the superpotential
\bea
W= \frac{1}{k} \sum_{\alpha=1}^n (-1)^\alpha \Tr \left( Q_\alpha Q_{\alpha+1} \tQ_{\alpha+1} \tQ_\alpha \right) \, , \label{superpot}
\eea
where we identify
\bea
Q_{n+1} := Q_{n} \, , \qquad \tQ_{n+1} := \tQ_{n} \, .
\eea

\subsubsection{A solution to the system of BAEs} \label{altCSN3}
Let us denote respectively by $\Delta_{\alpha}$, $\tilde{\Delta}_{\alpha}$ the chemical potentials associated to the flavor symmetries of $Q_\alpha$ and $\tQ_\alpha$, and by $n_{\alpha}$, $\tilde{n}_{\alpha}$ the fluxes associated with the flavor symmetries of $Q_\alpha$ and $\tQ_\alpha$.

From \eref{genbethepot}, the Bethe potential $\CV$ can be written as
{\small
\bea \label{potaltCS}
\frac{\CV}{i N^{3/2}} &= k \int dt \, t\, \rho(t) \sum_{\alpha=1}^m \delta v^{2\alpha-1}(t)
+ \int dt \, \rho(t)^2  \sum_{\alpha=1}^n \left[ g_+ \big(\delta v^\alpha(t)+ \tilde{\Delta}_{\alpha} \big) - g_- \big(\delta v^\alpha(t) - \Delta_{\alpha} \big) \right]  \nn \\
& - \frac{i}{N^{1/2}} \int dt\, \rho(t) \sum_{\alpha=1}^m \bigg[ \Li_2 \left(e^{i \big(\delta v^{2\alpha-1}(t)
+ \tilde{\Delta}_{2\alpha-1}\big)}\right) - \Li_2 \left(e^{i \big(\delta v^{2\alpha-1}(t)- {\Delta}_{2\alpha-1}\big)}\right) \nn \\
& + \Li_2 \left(e^{i \big(\delta v^{2\alpha}(t)+ \tilde{\Delta}_{2\alpha}\big)}\right)
- \Li_2 \left(e^{i \big(\delta v^{2\alpha}(t)- {\Delta}_{2\alpha}\big)}\right) \bigg] -\mu \left(\int dt \, \rho(t)-1 \right) \, ,
\eea}
where $\delta v^{\alpha}(t) = v^{\alpha+1}(t) - v^{\alpha}(t)$ and hence,
\bea
\sum_{\alpha=1}^n \delta v^{\alpha}(t) =0\, .
\eea
Without loss of generality, we set the chemical potentials associated with topological symmetries to zero. The subleading terms in \eref{potaltCS} can be obtained by considering the node $2 \alpha-1$ (with $\alpha=1, \ldots, m$), where the fields with chemical potentials $\tilde{\Delta}_{2\alpha-1}$, $\Delta_{2\alpha-2}$ are incoming to that node and those with chemical potentials ${\Delta}_{2\alpha-1}$, $\tilde{\Delta}_{2\alpha-2}$ are outgoing of that node.
This explains the signs of such terms in \eref{potaltCS}.
These terms can be neglected when we compute the value of the Bethe potential, since $\Li_2$ does not have divergences;
however, they play an important role when we deal with the derivatives of $\cV$ because $\Li_1(e^{iu})$ diverges as $u \to 0$.

Taking the variational derivatives of $\CV$ with respect to $\rho(t)$ and setting it to zero, we obtain
\bea \label{eq0}
0 &=k t \sum_{\alpha=1}^m \delta v^{2\alpha-1}(t) + 2 \rho(t) \sum_{\alpha=1}^n \left[ g_+ \big(\delta v^\alpha(t)+ \tilde{\Delta}_{\alpha}\big) - g_- \big(\delta v^\alpha(t) - \Delta_{\alpha}\big)  \right] - \mu \, .
\eea
When $\delta v^\alpha \not\approx - \tilde{\Delta}_\alpha$ and $\delta v^\alpha \not\approx \Delta_\alpha$ for all $\alpha$, setting the variational derivatives of $\CV$ with respect to $\delta v^\alpha(t)$ to zero yields
\bea \label{eqmid1}
0&= (-1)^{\alpha+1} k t + \rho(t) \Big[ g'_+\big( \delta v^\alpha(t) + \tilde{\Delta}_{\alpha}\big) - g'_-\big(\delta v^\alpha(t) - \Delta_{\alpha}\big) \nn \\
& + g'_-\big(\delta v^{\alpha-1}(t) - \Delta_{{\alpha-1}}\big) - g'_+\big(\delta v^{\alpha-1}(t) + \tilde{\Delta}_{{\alpha-1}}\big)  \Big]\, , \qquad \alpha =1, \ldots, n \, .
\eea
However, in the following, we also need to consider the cases in which  $\delta v^{2\alpha-1}(t) \approx -  \tilde{\Delta}_{2\alpha-1}$ and that in which $\delta v^{2\alpha-1}(t) \approx \Delta_{2\alpha-1}$, for all $\alpha =1, \ldots, m$.  
\bi
\item In the former case, taking $\delta v^{2\alpha-1}(t) = -  \tilde{\Delta}_{2\alpha-1}+\exp(- N^{1/2} \tilde{Y}_{2\alpha-1})$ and setting to zero the variational derivatives of $\CV$ with respect to $\delta v^{2\alpha-1}(t)$ and $\delta v^{2\alpha}(t)$ yields
\bea \label{eqleft1}
0&=\tilde{Y}_{2\alpha-1}(t) + k t + \rho(t) \Big[ g'_+(0) - g'_-\big( -\tilde{\Delta}_{2\alpha-1} - \Delta_{2\alpha-1}\big)\nn \\
& + g'_-\big(\delta v^{2\alpha-2}(t) - \Delta_{{2\alpha-2}}\big) - g'_+\big(\delta v^{2\alpha-2}(t) + \tilde{\Delta}_{2\alpha-2}\big)  \Big]\, , \nn \\
0&=-\tilde{Y}_{2\alpha-1} (t)-k t + \rho(t) \Big[ g'_+\big(\delta v^{2\alpha}(t) + \tilde{\Delta}_{2\alpha}\big) - g'_-\big(\delta v^{2\alpha}(t) - \Delta_{2\alpha}\big) \nn \\
& + g'_-\big(- \tilde{\Delta}_{2\alpha-1} - \Delta_{{2\alpha-1}}\big) - g'_+(0)  \Big]\, .
\eea
\item In the latter case, taking $\delta v^{2\alpha-1}(t) ={\Delta}_{2\alpha-1}-\exp(- N^{1/2} {Y}_{2\alpha-1})$ and setting to zero the variational derivatives of $\CV$ with respect to $\delta v^{2\alpha-1}(t)$ and $\delta v^{2\alpha}(t)$  yields
\bea \label{eqright1}
0&=-{Y}_{2\alpha-1}(t) + k t + \rho(t) \Big[ g'_+\big({\Delta}_{2\alpha-1} + \tilde{\Delta}_{{2\alpha-1}}\big) - g'_-(0) \nn \\
& + g'_-\big(\delta v^{2\alpha-2}(t) - \Delta_{{2\alpha-2}}\big) - g'_+\big(\delta v^{2\alpha-2}(t) + \tilde{\Delta}_{2\alpha-2}\big)  \Big]\, , \nn \\
0&={Y}_{2\alpha-1}(t) - k t + \rho(t) \Big[ g'_+\big(\delta v^{2\alpha}(t) + \tilde{\Delta}_{2\alpha}\big) - g'_-\big( \delta v^{2\alpha}(t) - \Delta_{2\alpha}\big) \nn \\
& + g'_-(0) - g'_+\big({\Delta}_{2\alpha-1} + \tilde{\Delta}_{{2\alpha-1}}\big)  \Big]\, .
\eea
\ei
We also impose the condition that the sum of the chemical potential for each term in the superpotential \eqref{superpot} is $2 \pi$,
\bea
\Delta_\alpha+\Delta_{\alpha+1}+\tilde{\Delta}_\alpha+\tilde{\Delta}_{\alpha+1} = 2\pi \, .
\eea

For later convenience, we define the following notations
\bea \label{paraA2m}
F_1 = m \sum_{\alpha=1}^m \Delta_{2\alpha} \, , \qquad
F_2 = m  \sum_{\alpha=1}^m \Delta_{2\alpha-1} \, , \qquad
F_3= \Delta_1 + \tilde{\Delta}_1 \, .
\eea

Let us now proceed to solve the BAEs. First, we solve \eref{eq0}-\eref{eqmid1} and obtain
\be
\begin{aligned}
\rho & =  \frac{m k t\left[F_1 F_3-F_2 (2\pi -F_3)\right] +2 \pi  \mu }{m F_3 (2 \pi -F_3 ) \left(2 \pi-F_1-F_2 \right) \left(F_1+F_2\right) } \\[.5em]
\delta v^{2\alpha-1} &= \Delta_{2\alpha-1} -\frac{\left(F_1+F_2\right) F_3 \left[ \mu -m k t(2 \pi -F_3-F_1) \right]}{m k t \left[F_1 F_3-F_2 (2\pi -F_3)\right]+2 \pi  \mu } \\[.5em]
\delta v^{2\alpha} &= \Delta_{2\alpha} -\frac{\left(F_1+F_2\right) (2\pi -F_3) \left[ \mu +m k t(F_3-F_2) \right]}{m k t \left[F_1 F_3-F_2 (2\pi -F_3)\right]+2 \pi  \mu }
\end{aligned}
\qquad\qquad t_< < t < t_> \, .
\ee
This solution is valid in the interval $[t_{<},t_>]$ where the end points are determined from 
\bea
\delta v^{2\alpha-1}(t_<) = - \tilde{\Delta}_{2\alpha-1} \, , \qquad \delta v^{2\alpha-1}(t_>) = \Delta_{2\alpha-1}  \quad \text{for all $\alpha=1, \ldots, m$} \, .
\eea 
Explicitly, they are
\bea
t_< = - \frac{\mu}{k m F_1}\, ,\qquad 
t_> = \frac\mu{k m (2 \pi - F_1 -F_3)}\, .
\eea
Next, we focus on the regions $[t_{\ll}, t_<] $ and $[t_>, t_{\gg}]$, where $\delta v^{2\alpha-1}(t) = -  \tilde{\Delta}_{2\alpha-1}$ for $ t \in [t_{\ll}, t_<]$ and $\delta v^{2\alpha-1}(t) = {\Delta}_{2\alpha-1}$ for $ t \in [t_{>}, t_\gg]$.

For the interval $[t_{\ll}, t_<] $, we solve \eref{eq0} and \eref{eqleft1} and obtain
\be
\begin{aligned}
\rho & = \frac{\mu +m \left(F_3 -F_3  \right) k t}{m F_3 \left(F_1+F_2-F_3\right) \left(2  \pi -F_1-F_2\right)} \\[.5em]
\delta v^{2\alpha-1} &= -  \tilde{\Delta}_{2\alpha-1} \, ,\qquad  \delta v^{2\alpha} = F_3-F_1-F_2 + \Delta_{2\alpha} \\[.5em]
\widetilde{Y}_{2\alpha-1}&=  -\frac{\mu +m kt F_1 }{m(F_3-F_1-F_2)}
\end{aligned}
\qquad\qquad t_\ll < t < t_< \, ,
\ee
where we determine the end point $t_\ll$ by the condition $\rho(t_\ll)=0$:
\bea
t_\ll = - \frac{\mu}{km( F_3-F_2)}
\eea

For the interval $[t_>, t_\gg]$, we solve \eref{eq0} and \eref{eqright1} and obtain
\be
\begin{aligned}
\rho & =  \frac{\mu-m k t F_2}{m F_3 \left(F_1+F_2-F_3\right) \left(2\pi- F_1-F_2\right)} \\[.5em]
\delta v^{2\alpha-1} &= \Delta_{2\alpha-1}\, , \qquad \delta v^{2\alpha} =  -F_1-F_2 + \Delta_{2\alpha} \\[.5em]
{Y}_{2\alpha-1} &=  \frac{\mu-m kt(2\pi- F_1-F_3)}{m (2\pi - F_1 - F_2-F_3)}
\end{aligned}
\qquad\qquad t_> < t < t_\gg \, ,
\ee
where we determine the end point $t_\gg$ by the condition $\rho(t_\gg)=0$:
\bea
t_\gg = \frac\mu{k m F_2}\, .
\eea

To summarize, the above solution is divided into three regions, namely the {\it left tail} $[t_\ll, t_<]$, the {\it inner interval} $[t_<, t_>]$ and the {\it right tail} $[t_>, t_\gg]$.  These are depicted in the following diagram:
\begin{center}
\begin{tikzpicture}[scale=2]
\draw (-2.5,0) -- (2.5,0);
\draw (-2.5,-.05) -- (-2.5, .05); \draw (-1,-.05) -- (-1, .05); \draw (1,-.05) -- (1, .05); \draw (2.5,-.05) -- (2.5, .05);
\node [below] at (-2.5,0) {$t_\ll$}; \node [below] at (-2.5,-.3) {$\rho=0$};
\node [below] at (-1.,0) {$t_<$}; \node [below] at (-1.,-.3) { $\delta v^{2\alpha-1} = - \tilde{\Delta}_{2\alpha-1} \, \forall \alpha$}; %\node [below] at (-0.5,-.6) {\footnotesize$\widetilde{Y}_{2\alpha-1} = 0$};
\node [below] at (1,0) {$t_>$}; \node [below] at (1,-.3) { $\delta v^{2\alpha-1}=\Delta_{2\alpha-1}  \, \forall \alpha $}; 
%\node [below] at (0.5,-.6) {\quad \footnotesize$Y_{2\alpha-1} = 0$};
\node [below] at (2.5,0) {$t_\gg$}; \node [below] at (2.5,-.3) {$\rho=0$};
\end{tikzpicture}
\end{center}
Finally, the normalization $\int_{t_\ll}^{t_\gg} dt \, \rho(t) =1$ fixes
\be
\label{solution sum 2pi -- end -- A2m}
\mu =m \sqrt{2k F_1  F_2 \left(F_3-F_2\right) \left(2 \pi -F_3- F_1 \right)} \, .
\ee

\subsubsection{The index at large $N$}

From \eref{genentropy}, the topological free energy of this theory is given by
\bea
\frac{\mathfrak{F}}{N^{3/2}} &= -\int dt \, \rho(t)^2 \Bigg\{ \frac{n \pi^2}{3} + \sum_{\alpha=1}^n \left[ (\tilde{n}_{\alpha} -1) g'_+(\delta v^\alpha+ \tilde{\Delta}_{\alpha})
+ (n_{\alpha} -1) g'_- (\delta v^\alpha - \Delta_{\alpha})\right]\Bigg\} \nn \\
& \qquad - \sum_{\alpha=1}^m \tilde{n}_\alpha \int_{\delta v^{2\alpha-1} \approx -\tilde{\Delta}_{2\alpha-1}} dt \, \rho(t) \, \tilde{Y}_{2\alpha-1}(t)   - \sum_{\alpha=1}^m n_\alpha \int_{\delta v^{2\alpha-1} \approx {\Delta}_{2\alpha-1}}  dt \, \rho(t) \, {Y}_{2\alpha-1}(t) \, .
\eea
The result depends only on the parameters $F_1$, $F_2$, $F_3$ and their corresponding flavor magnetic fluxes
\bea
{\frak n}_1 = m \sum_{\alpha=1}^m n_{2\alpha} \, , \qquad {\frak n}_2 = m \sum_{\alpha=1}^m n_{2\alpha-1} \, , \qquad {\frak n}_3 = n_1 + \tilde{n}_1 \, ,
\eea
and can be written as
\bea \label{freealtCS}
\mathfrak{F} = m\, \mathfrak{F}_{\text{ABJM}_k} \, .
\eea
The map of the parameters is as follows,
\bea
\Delta_{A_1} & = F_1 \, , \qquad \Delta_{A_2} =  F_2 \, , \qquad \Delta_{B_1}=F_3-F_2\, , \qquad \Delta_{B_2} = 2\pi -F_1- F_3 \, , \nn \\
\fn_{A_1} & =  {\frak n}_1 \, , \qquad \fn_{A_2} = {\frak n}_2 \, , \qquad \fn_{B_1}= {\frak n}_3-{\frak n}_2 \, , \quad \fn_{B_2} = 2 - {\frak n}_1- {\frak n}_3 \, .
\eea

Recall that the geometric branch of the moduli space of this theory is $\Sym^N (\BC^2/\BZ_m \times \BC^2/\BZ_{m})/\BZ_k$, whereas that of the ABJM theory is $\Sym^N (\BC^4/\BZ_k)$.
The square root of the relative orbifold orders of these two spaces explains the prefactor $m$ in \eref{freealtCS}.

\subsection{The affine $A_{n-1}$ quiver with two adjacent CS levels of opposite signs} \label{sec:necklacetwoCS}
We are interested in the necklace quiver with $n$ nodes, each with $\U(N)$ gauge group, and the Chern-Simons levels:
\bea
k_\alpha= \begin{cases} +k & \text{if $\alpha=1$} \\
-k & \text{if $\alpha=2$}  \\
0 & \text{otherwise}
\end{cases}
\eea 
The $\CN=2$ quiver diagram of this theory is
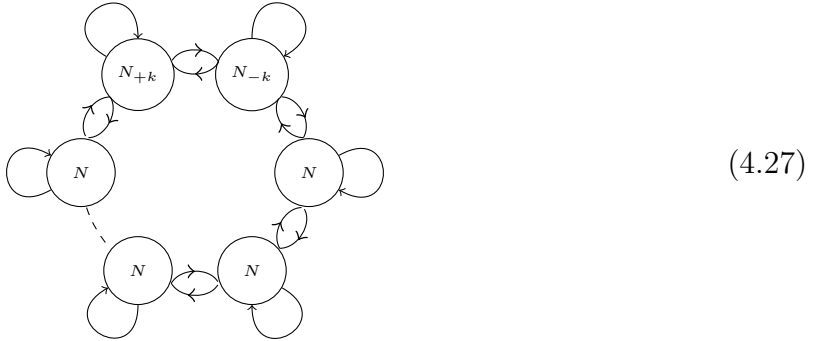
\bea \label{fig:CStwo}
\begin{tikzpicture}[baseline,font=\tiny, every loop/.style={min distance=10mm,looseness=10}, every node/.style={minimum size=0.9cm}]
\def \n {6}
\def \radius {1.5cm}
\def \margin {18} % margin in angles, depends on the radius
\foreach \s in {1}
{
    \node[draw, circle] at ({2*360/\n * (\s)}:\radius)  (\s) {$N_{+k}$};
    \draw[black,-> ] (\s) edge [out={30+2*\s*60},in={-30+2*\s*60},loop,looseness=5] (\s);
} 
\foreach \s in {1}
{    
    \node[draw, circle]  at ({2*360/\n*\s-360/\n}:\radius) (\s)  {$N_{-k}$};
    \draw[black,-> ] (\s) edge [out={-30+2*\s*60},in={-90+2*\s*60},loop,looseness=5] (\s);
} 
\foreach \s in {2,3}
{
    \node[draw, circle] at ({2*360/\n * (\s)}:\radius)  (\s) {$N$};
    \draw[black,-> ] (\s) edge [out={30+2*\s*60},in={-30+2*\s*60},loop,looseness=5] (\s);
} 
\foreach \s in {2,3}
{    
    \node[draw, circle]  at ({2*360/\n*\s-360/\n}:\radius) (\s)  {$N$};
    \draw[black,-> ] (\s) edge [out={-30+2*\s*60},in={-90+2*\s*60},loop,looseness=5] (\s);
} 
\foreach \s in {1,...,5}
{  
  \draw[decoration={markings, mark=at position 0.5 with {\arrow[scale=1.5]{>}}}, postaction={decorate}, shorten >=0.7pt]  ({360/\n * (\s - 3)+\margin}:\radius) to[bend left=60] node[midway,above] {}node[midway,below] {}  ({360/\n * (\s - 2)-\margin}:\radius) ; 
    \draw[decoration={markings, mark=at position 0.5 with {\arrow[scale=1.5]{<}}}, postaction={decorate}, shorten >=0.7pt]  ({360/\n * (\s - 3)+\margin}:\radius) to[bend right=60] node[midway,above] {}node[midway,below] {}  ({360/\n * (\s - 2)-\margin}:\radius) ;
}   
\draw[dashed, >=latex] ({360/6 * (5 -2)+\margin}:\radius) 
    arc ({360/6 * (5 -2)+\margin}:{360/6 * (5-1)-\margin}:\radius);
%\node[draw=none] at (5,-1.5) {{\footnotesize ($\bullet$ = $U(N)$ gauge node; $n$ gauge nodes)}};
\end{tikzpicture}
\eea
In the notation of the preceding subsection, the superpotential can be written as
\bea
W = \sum_{\alpha=1}^n  \Tr \left(Q_\alpha \phi_{\alpha+1} \tilde{Q}_\alpha- \tilde{Q}_{\alpha} \phi_{\alpha} Q_{\alpha}\right) + \frac{k}{2} \Tr \left(\phi_1^2 - \phi_2^2\right) \, .
\eea
After integrating out the massive adjoint fields $\phi_1$ and $\phi_2$, we have the superpotential
\bea
W &=-\frac{1}{k} \Tr \left(Q_1 Q_2 \tQ_2 \tQ_1  - Q_1 \tilde{Q}_1 \tilde{Q}_n Q_n \right)+ \frac{1}{2k} \Tr \left[\left(Q_2 \tilde{Q}_2\right)^2 - \left(Q_n \tilde{Q}_n \right)^2 \right] \nn \\
&+ \sum_{\alpha=2}^{n-1} \Tr \left(Q_\alpha \phi_{\alpha+1} \tilde{Q}_\alpha - \tilde{Q}_{\alpha+1} \phi_{\alpha+1} Q_{\alpha+1} \right), \label{superpot1}
\eea

\subsubsection{A solution to the system of BAEs}
Let us denote respectively by $\Delta_{\alpha}$, $\tilde{\Delta}_{\alpha}$, $\Delta_{\phi_\alpha}$ the chemical potentials associated to the flavor symmetries of $Q_\alpha$, $\tQ_\alpha$, $\phi_\alpha$, and by $n_{\alpha}$, $\tilde{n}_{\alpha}$, ${n}_{\phi_\alpha}$ the corresponding fluxes associated with their flavor symmetries.  We also denote by $\Delta_m^{(\alpha)}$ the chemical potential associated with the topological charge corresponding to node $\alpha$ and $\ft^{(\alpha)}$ the corresponding magnetic flux.

The superpotential \eqref{superpot1} implies the following constraints
\bea
&\tilde{\Delta}_\alpha = \pi - \Delta_\alpha \quad \text{for all $\alpha=1,\ldots,n$} \nn \\
&\Delta_{\phi_3} = \ldots =  \Delta_{\phi_n} = \pi \, . 
\eea

The Bethe potential for this particular model can be derived from formula \eref{genbethepot}.
The procedure of solving the BAEs  is similar to that presented in section \ref{altCSN3}.
The solution can be separated into three regions, namely the left tail $[t_{\ll}, t_<]$, the inner interval $[t_{<},t_>]$ and the right tail $[t_>, t_{\gg}]$, where
\bea
t_< \text{ s.t. } \delta v^{1}(t_<) = -  \tilde{\Delta}_{1} \, , \qquad
 t_> \text{ s.t. } \delta v^{1}(t_>) = \Delta_{1} \, .
\eea
The end points $t_\ll$ and $t_\gg$ are the values where $\rho=0$ on the left and the right tails, respectively. Schematically:
\begin{center}
\begin{tikzpicture}[scale=2]
\draw (-2,0) -- (2,0);
\draw (-2.,-.05) -- (-2., .05); \draw (-0.7,-.05) -- (-0.7, .05); \draw (0.7,-.05) -- (0.7, .05); \draw (2,-.05) -- (2, .05);
\node [below] at (-2.,0) {$t_\ll$}; \node [below] at (-2.,-.3) {$\rho=0$};
\node [below] at (-0.7,0) {$t_<$}; \node [below] at (-0.7,-.3) {$\delta v^{1} = - \tilde{\Delta}_{1}$}; %\node [below] at (-0.5,-.6) {\footnotesize$\widetilde{Y}_{2\alpha-1} = 0$};
\node [below] at (0.7,0) {$t_>$}; \node [below] at (0.7,-.3) { $\delta v^{1}=\Delta_{1}$}; 
%\node [below] at (0.5,-.6) {\quad \footnotesize$Y_{2\alpha-1} = 0$};
\node [below] at (2,0) {$t_\gg$}; \node [below] at (2,-.3) {$\rho=0$};
\end{tikzpicture}
\end{center}

It turns out that the solution depends on the following parameters:
\bea \label{paraF}
F_1 = \Delta_1 + \frac{1}{k} \sum_{\alpha=1}^n \Delta_m^{(\alpha)} \, , \qquad
F_2 = \frac{1}{n-1} \left[  \left( \sum_{\alpha=2}^n \Delta_{\alpha}  \right) - \frac{1}{k} \sum_{\alpha=1}^n \Delta_m^{(\alpha)} \right] \, .
\eea 

The solution is as follows.
In the left tail $[t_\ll, t_<]$, we have
\be
\begin{aligned}
\rho & =  \frac{(n-1) \left[ \mu +(\pi -F_1 )k t\right]}{\pi  \left[n\pi -F_1-(n-1)F_2 \right] \left[\pi -F_1-(n-1)F_2\right]} \\[.5em]
\delta v^{1} &= -  \tilde{\Delta}_{1}  \\[.5em]
\delta v^{\alpha} &=\Delta_\alpha + \left[ \pi-F_1-(n-1)F_2 \right] \, , \qquad \forall \, 2 \leq \alpha \leq n  \\[.5em]
\widetilde{Y}_{1} &=  \frac{(n-1)F_2k t+\mu }{\pi-F_1-(n-1)F_2 } \, .
\end{aligned}
\ee
In the inner interval $[t_<, t_>]$, we have
\be
\begin{aligned}
\rho & =  \frac{\left[(n-1) (F_1-F_2)\right] k t-n \mu }{\pi  \left[F_1+(n-1)F_2\right] \left[F_1+(n-1)F_2-n \pi \right]} \\[.5em]
\delta v^{1} &= \frac{ \mu  n \Xi-(n-1) \Big[ F_1 [ \mu + (\pi+\Xi  )k t ]-F_2 [\mu- \{ (n-1)\pi-\Xi \} k t ] \Big]+(n-1) [F_1^2 +(n-1)F_2^2] k t }{(n-1)(F_1-F_2) k t- n \mu } \\[.5em]
\delta v^{\alpha} &= \frac{\left[F_1+(n-1)F_2\right] \left[\mu +\left(\pi -F_1 \right) k t\right]}{\left[(n-1) F_1-(n-1)F_2\right] k t-n \mu } +\Delta_\alpha~, \quad \forall \, 2\leq \alpha \leq n \, ,
\end{aligned}
\ee
where $\Xi = \frac{1}{k} \sum_\alpha \Delta_m^{(\alpha)}$.
In the right tail $[t_>, t_\gg]$ we have
\be
\begin{aligned}
\rho &=  \frac{(n-1) \left(F_1 k t-\mu \right)}{\pi  \left[F_1+(n-1)F_2\right] \left[F_1+(n-1)F_2-(n-1) \pi \right]} \\[.5em]
\delta v^{1} &=  {\Delta}_{1}  \\[.5em]
\delta v^{\alpha} &= \Delta_\alpha +\frac{1}{n-1} \left[  \pi -F_1-(n-1)F_2 \right] \\[.5em]
{Y}_{1} &=  \frac{\mu - (n-1)(\pi-F_2) k t}{F_1+(n-1)F_2-(n-1) \pi } \, .
\end{aligned}
\ee
The transition points are at
\bea \label{solution sum 2pi -- init -- An}
t_\ll = - \frac{\mu }{k (\pi - F_1)} \, ,\qquad t_< = - \frac{\mu }{k F_2} \, ,\qquad
t_> = \frac{\mu }{k(n-1) (\pi-F_2)} \, ,\qquad t_\gg = \frac{\mu }{k  F_1} \, .
\eea
Finally, the normalization $\int_{t_\ll}^{t_\gg} dt \, \rho(t) =1$ fixes
\be
\mu = \sqrt{2(n-1)k  F_1 F_2 (\pi-F_1) ( \pi - F_2) } \, .
\ee

\subsubsection{The index at large $N$}
The topological free energy of this theory can be derived from \eref{genentropy}.
We find that the topological free energy of this quiver theory depends only on the parameters $F_1$, $F_2$ given by \eref{paraF} and their corresponding conjugate charges
\bea
{\frak n}_1 = n_1 + \frac{1}{k} \sum_{\alpha=1}^n \ft_\alpha\, , \qquad  {\frak n}_2 =\frac{1}{n-1} \left[ \left( \sum_{\alpha=2}^n n_{\alpha} \right)-  \frac{1}{k} \sum_{\alpha=1}^n \ft_\alpha \right]\, .
\eea
The topological free energy can be written as,
\bea
\mathfrak{F}= \sqrt{n-1}\, \mathfrak{F}_{\text{ABJM}}\, . \label{freetwoaltCS}
\eea
The map of the parameters is as follows,
\bea
\Delta_{A_1} & = F_1\, , \qquad \Delta_{A_2} = F_2\, , \qquad \Delta_{B_1}= \pi - F_1\, , \qquad \Delta_{B_2} = \pi - F_2\, , \nn \\
\fn_{A_1} & =  \fn_1\, , \qquad \fn_{A_2} =\fn_2\, , \qquad \fn_{B_1}= 1 - \fn_1\, , \qquad \fn_{B_2} =1 - \fn_2\, .
\eea
Indeed, for $n=2$, this theory becomes the ABJM theory and \eref{freetwoaltCS} reduces to $\mathfrak{F}_{\text{ABJM}}$, as expected.
Recall that the geometric branch of the moduli space of this theory is $\Sym^N (\BC^2 \times \BC^2/\BZ_{n-1})/\BZ_k$, whereas that of the ABJM theory is $\Sym^N (\BC^4/\BZ_k)$.
The square root of the relative orbifold orders of these two spaces explains the prefactor $\sqrt{n-1}$ in \eref{freetwoaltCS}.

Let us also comment on the number of the parameters which appears in the topological free energy of this model.
It can be seen from \eref{freetwoaltCS} that the topological free energy depends only on two parameters, $F_1$ and $F_2$ (or $\fn_1$ and $\fn_2$), instead of three, despite the fact that the geometric branch is associated with Calabi-Yau four-fold $\BC^2 \times \BC^2/\BZ_{n-1}$.
Indeed, in the $\CN=3$ description of the quiver, only $\U(1)^2$ (one mesonic and one topological symmetry) is manifest (see Appendix C of \cite{Cremonesi:2016nbo}).
An extra mesonic symmetry that exchanges the holomorphic variables on $\BC^2$ and those on $\BC^2/\BZ_2$ is not present in the quiver description of this theory.

\paragraph{$\SL(2,\BZ)$ duality.} The affine $A_{n-1}$ quiver \eref{fig:CStwo} with $n$ gauge nodes and $k=1$ is $\SL(2,\BZ)$ dual to the $A_{n-2}$ Kronheimer-Nakajima quiver \eref{KNN4quiv} with $n-1$ gauge nodes and $r=1$.
This duality can be seen from the Type IIB brane configuration as follows \cite{Hanany:1996ie, Aharony:1997ju, Gaiotto:2008ak, Assel:2014awa}.
The configuration of the Kronheimer-Nakajima quiver involves $N$ D3-branes wrapping $\BR^{1,2}_{0,1,2} \times S^{1}_6$ (where the subscripts indicate the direction in $\BR^{1,9}$);
$n-1$ NS5-branes wrapping $\BR^{1,2}_{0,1,2} \times \BR^3_{7,8,9}$ located at different positions along the circular $x^6$ direction;
and $r=1$ D5-branes wrapping $\BR^{1,2}_{0,1,2} \times \BR^3_{3,4,5}$ located along the circular $x^6$ direction within one of the NS5-brane intervals.
Applying an $\SL(2,\BZ)$ action on such a configuration, we can obtain a similar configuration except that the D5-brane becomes a $(1,1)\,5$-brane.
This is in fact the configuration for quiver \eref{fig:CStwo} with $n$ gauge nodes and $k=1$.
Indeed, in this case we can match the topological free energies \eref{freetwoaltCS} and \eref{freeKN}, as expected from the duality.

\subsection{The $N^{0,1,0}/\BZ_k$ theory} \label{sec:N010}
In this section we focus on the holographic dual of M-theory on ${\rm AdS}_4 \times N^{0,1,0}/\BZ_k$ \cite{Fabbri:1999hw, Billo:2000zr, Yee:2006ba}.
$N^{0,1,0}$ is a homogeneous Sasakian of dimension seven and defined as the coset $\SU(3)/\U(1)$. The manifold has the isometry $\SU(3) \times \SU(2)$.
The latter $\SU(2)$ is identified with the R-symmetry.
The description of the dual field theory was discussed in \cite{Gaiotto:2009tk, Imamura:2011uj, Cheon:2011th}.
This theory has $\CN=3$ supersymmetry and contains $\CG = \U(N)_{+k} \times \U(N)_{-k}$ gauge group with two bi-fundamental hypermultiplets and $r$ flavors of fundamental hypermultiplets under one of the gauge groups.
The $\CN=3$ quiver is depicted as follows:
\bea
\begin{tikzpicture}[font=\footnotesize, scale=0.9]
\begin{scope}[auto,%
  every node/.style={draw, minimum size=0.5cm}, node distance=2cm];
\node[circle]  (UN1)  at (0.3,1.7) {$N_{+k}$};
\node[circle, right=of UN1] (UN2) {$N_{-k}$};
\node[rectangle, left=of UN1] (Ur) {$r$};
\end{scope}
\draw[draw=blue,solid,line width=0.1mm]  (UN1) to[bend right=30]  (UN2) ; 
\draw[draw=red,solid,line width=0.1mm]  (UN1) to[bend left=30]  (UN2) ;    
\draw[draw=black,solid,line width=0.1mm]  (UN1) to (Ur) ;    
\end{tikzpicture}
\eea
Note that for $k=0$, this theory becomes the Kronheimer-Nakajima quiver \eref{KNN4quiv} with $n=2$.

In $\CN=2$ notation, the quiver diagram for this theory is
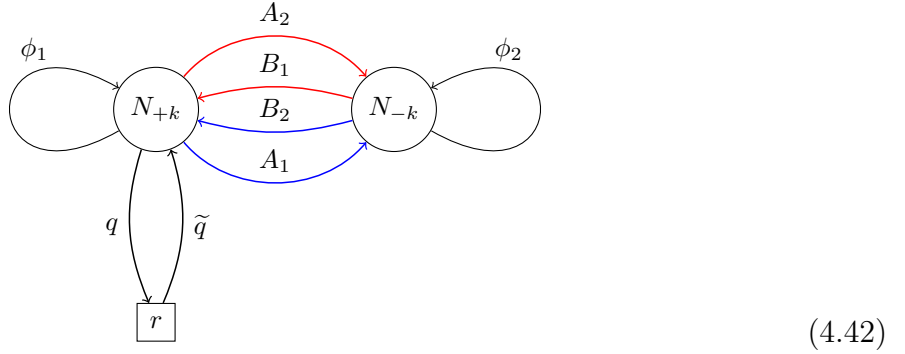
\bea
\begin{tikzpicture}[font=\footnotesize, scale=0.8]
\begin{scope}[auto,%
  every node/.style={draw, minimum size=0.5cm}, node distance=2cm];
  % the vertices
\node[circle] (USp2k) at (0., 0) {$N_{+k}$};
\node[circle, right=of USp2k] (BN)  {$N_{-k}$};
\node[rectangle, below=of USp2k] (Ur)  {$r$};
\end{scope}
  % the edges
\draw[draw=blue,solid,line width=0.2mm,<-]  (USp2k) to[bend right=15] node[midway,above] {$B_2 $}node[midway,above] {}  (BN) ;
\draw[draw=blue,solid,line width=0.2mm,->]  (USp2k) to[bend right=50] node[midway,above] {$A_1$}node[midway,above] {}  (BN) ; 
\draw[draw=red,solid,line width=0.2mm,<-]  (USp2k) to[bend left=15] node[midway,above] {$B_1$} node[midway,above] {} (BN) ;  
\draw[draw=red,solid,line width=0.2mm,->]  (USp2k) to[bend left=50] node[midway,above] {$A_2$} node[midway,above] {} (BN) ;    
\draw[black,-> ] (USp2k) edge [out={-150},in={150},loop,looseness=10] (USp2k) node at (-2.,1) {$\phi_1$} ;
\draw[black,-> ] (BN) edge [out={-30},in={30},loop,looseness=10] (BN) node at (5.8,1) {$\phi_2$};
%\draw (0,0.55) arc (4:280:0.75cm) node at (-2,1) {$\phi_1$} ;
%\draw(4,0.55) arc (0:-300:-0.75cm) node at (5.8,1) {$\phi_2$} ;
\draw[draw=black,solid,line width=0.2mm,<-]  (USp2k) to[bend left=20] node[midway,right] {$\tilde{q}$} node[midway,above] {} (Ur) ;  
\draw[draw=black,solid,line width=0.2mm,->]  (USp2k) to[bend right=20] node[midway,left] {$q$} node[midway,above] {} (Ur) ;    
\end{tikzpicture}
\eea
where the bi-fundamental chiral fields $(A_1, B_2)$ come from one of the $\CN=3$ hypermultiplet indicated in blue, and the bi-fundamental chiral fields $(A_2, B_1)$ come from the other $\CN=3$ hypermultiplet indicated in red. The superpotential is given by
\bea \label{WN010mass}
W = \Tr \left(A_1 \phi_2 B_2 - B_2 \phi_1 A_1 - A_2 \phi_2 B_1 + B_1 \phi_1 A_2 + \frac{k}{2} \phi_1^2 - \frac{k}{2} \phi_2^2 + \tilde{q} \phi_1 q \right) \, .
\eea
Note that the bi-fundamental fields $A_1, A_2, B_1, B_2$ can be mapped to those in the Kronheimer-Nakajima quiver \eref{KNN2quiv} with $n=2$ as follows
\bea
A_1 \, \leftrightarrow \, Q_1\, , \qquad A_2 \, \leftrightarrow \, \tQ_2 \, , \qquad B_1 \, \leftrightarrow \, Q_2\, , \qquad B_2 \, \leftrightarrow \, \tQ_1 \, .
\eea

Integrating out the massive adjoint fields $\phi_{1,2}$ in \eref{WN010mass}, we obtain the superpotential
\begin{equation}\label{superpotentialN010}
 W = \Tr\left[\left(\epsilon^{ij} B_i A_j - q \tilde q\right)^2-\left(\epsilon^{ij} A_i B_j\right)^2\right]\, .
\end{equation}

\subsubsection{A solution to the system of BAEs}
The Bethe potential for this particular model can be derived from formula \eref{genbethepot}.  The procedure of solving the BAEs  is similar to that presented in sections \ref{sec:solnKN} and \ref{altCSN3}.  In the following we present an explicit solution to the corresponding BAEs.

For brevity, let us write
\bea \label{shorthandAB}
\Delta_1 & = \Delta_{A_1}\, , \qquad \Delta_2 = \Delta_{A_2}\, , \qquad \Delta_3 = \Delta_{B_1}\, , \qquad \Delta_4 = \Delta_{B_2} \, , \nn \\
\fn_1 & = \fn_{A_1}\, , \qquad \fn_2 = \fn_{A_2}\, ,\qquad \fn_3 = \fn_{B_1}\, , \qquad \fn_4 = \fn_{B_2}\, . 
\eea
We look for a solution to the BAEs such that
\begin{align}\label{Delta constraintN010}
\Delta_{q}+ \Delta_{\tilde{q}}=\pi \, , \qquad \Delta_{1} + \Delta_{4} = \pi \, , \qquad \Delta_{2} + \Delta_{3} = \pi \, ,
\end{align}
and
\begin{align}\label{flux constraintN010}
 &\fn_{q}+\fn_{\tilde{q}}=1\, ,\qquad\qquad
 \fn_{1} + \fn_{4} = 1\, ,\qquad\qquad
 \fn_{2} + \fn_{3} = 1\, .
\end{align}

The solution can be separated into three regions, namely the left tail $[t_{\ll}, t_<]$, the inner interval $[t_{<},t_>]$ and the right tail $[t_>, t_{\gg}]$, where
\be
t_< \text{ s.t. } \delta v(t_<) = - \Delta_3 \,,\qquad\qquad t_> \text{ s.t. } \delta v(t_>) = \Delta_1 \,.
\ee
Then we define $t_\ll$ and $t_\gg$ as the values where $\rho=0$ and those bound the left and right tails. Schematically:
\begin{center}
\begin{tikzpicture}[scale=2]
\draw (-1.5,0) -- (1.5,0);
\draw (-1.5,-.05) -- (-1.5, .05); \draw (-0.5,-.05) -- (-0.5, .05); \draw (0.5,-.05) -- (0.5, .05); \draw (1.5,-.05) -- (1.5, .05);
\node [below] at (-1.5,0) {$t_\ll$}; \node [below] at (-1.5,-.3) {$\rho=0$};
\node [below] at (-0.5,0) {$t_<$}; \node [below] at (-0.5,-.3) {$\delta v = -\Delta_3$};% \node [below] at (-0.5,-.6) {$Y_3 = 0$};
\node [below] at (0.5,0) {$t_>$}; \node [below] at (0.5,-.3) {$\delta v=\Delta_1$};% \node [below] at (0.5,-.6) {$Y_1 = 0$};
\node [below] at (1.5,0) {$t_\gg$}; \node [below] at (1.5,-.3) {$\rho=0$};
\end{tikzpicture}
\end{center}

The solution is as follows.
In the left tail we have
\be
\begin{aligned}
\rho &= \frac{\mu + k t\Delta_3 - \frac{\pi}{2}r  |t|}{\pi (\Delta_1 + \Delta_3) (\Delta_4 - \Delta_3)} \\[.5em]
\delta v &= - \Delta_3 \,,\qquad\qquad Y_3 = \frac{- kt\Delta_4 -\mu + \frac{\pi}{2}r |t|}{\Delta_4 - \Delta_3}
\end{aligned}
\qquad\qquad t_\ll < t < t_< \, .
\ee
In the inner interval we have
\be
\begin{aligned}
\rho &= \frac{2 \mu + k t(\Delta_3 - \Delta_1) - \pi r |t|}{\pi (\Delta_1 + \Delta_3)(\Delta_2 + \Delta_4)} \\[.5em]
\delta v &= \frac{\left( \mu - \frac{\pi}{2} r |t| \right) (\Delta_1 - \Delta_3) +k t \left( \Delta_1 \Delta_4 + \Delta_2 \Delta_3 \right)}{2 \mu + k t(\Delta_3 - \Delta_1) - \pi r |t|}
\end{aligned}
\qquad\qquad t_< < t < t_> \, ,
\ee
and $\delta v'>0$. In the right tail we have
\be
\begin{aligned}
\rho &= \frac{\mu -k t \Delta_1 - \frac{\pi}{2}r |t|}{\pi (\Delta_1 + \Delta_3)(\Delta_2 - \Delta_1)} \\[.5em]
\delta v &= \Delta_1 \,,\qquad\qquad Y_1 = \frac{kt\Delta_2 - \mu + \frac{\pi}{2}r  |t|}{\Delta_2 - \Delta_1}
\end{aligned}
\qquad\qquad t_> < t < t_\gg \, .
\ee
The transition points are at
\bea
\label{solution sum 2pi -- init -- N010}
& t_\ll = - \frac{2\mu}{\pi r + 2 k  \Delta_3} \, ,\qquad t_< = - \frac{2\mu}{\pi r + 2 k \Delta_4} \, ,\qquad
& t_> = \frac{2\mu}{\pi r + 2 k  \Delta_2} \, ,\qquad t_\gg = \frac{2\mu}{\pi r + 2 k  \Delta_1} \, .
\eea
Finally, the normalization fixes
{\small
\be
\label{solution sum 2pi -- end -- N010}
\mu = \frac12 \sqrt{\frac{\left[ \pi^2 - (\Delta _3-\Delta _4)^2 \right] \left[\pi  (2 k+r)-2 k\Delta _3\right] \left[\pi  (2 k+r)-2 k \Delta _4 \right] \left( 2k  \Delta _3 +\pi  r\right)  \left(2 k \Delta _4 +\pi  r\right)}{2 k^2 \left[2 k \Delta _3 \Delta _4+ \pi r (\Delta _3+\Delta _4) - (k+r) (\Delta _3^2+\Delta _4^2) \right]+\pi ^2 \left(2 k^3+4 k^2 r+4 k r^2+r^3\right)}} \, .
\ee}For $k=0$, this expression indeed reduces to \eref{muKN} with 
\bea
F_1 & = 2 \pi c\, , \qquad F_2 = c (\Delta_3 - \Delta_4 - \pi ) \, , \qquad F_3 = 2 \pi - c (\Delta_3 - \Delta_4 + \pi) \, , \nn \\
\Delta_m & = 2 \pi + c (\Delta_3 - \Delta_4 + \pi) \, ,
\eea
and $c =1/(2 \times 12^{1/3})$. Note that $F_1+F_2+F_3 = 2 \pi$, as required.
\subsubsection{The index at large $N$}
The topological free energy of this theory can be computed from \eref{genentropy}.
The expression for the topological free energy is fairly long, so we will just give the formul\ae{} for $k=1$, $r=1$ and
\begin{equation}
 \Delta_3 = \Delta_4 = \Delta \, , \qquad  \fn_3 = \fn_4 = \fn \, .
\end{equation}
In this case, the topological free energy reads
\begin{align}
\mathfrak{F} & = - \frac{2 N^{3/2}}{3} \frac{\pi (\pi - 2 \Delta ) \left[ 4 (\pi -\Delta ) \Delta + 19 \pi^2 \right] \fn
 + \left( 8 \Delta^4 - 20 \pi \Delta^3 - 6 \pi^2 \Delta^2 + 37 \pi^3 \Delta + 33 \pi^4 \right)}{\left[ 4 (\pi - \Delta ) \Delta + 11 \pi^2 \right]^{3/2}} \, .
\end{align}

\section{Quivers with $\CN=2$ supersymmetry} \label{sec:N2susy}
Let us now consider quiver gauge theories with $\CN=2$ supersymmetry. We first discuss non-toric theories associated with the Sasaki-Einstein seven manifold $V^{5,2}$.  There are two known models in this cases, one proposed by \cite{Martelli:2009ga} and the other by \cite{Jafferis:2009th}. We show that the topological free energy of these models can be matched with each other. We then move on to discuss flavored toric theories \cite{Benini:2009qs}. The procedure in solving the BAEs for these theories is similar to that for $\CN=3$ theories discussed in the preceding section.  

\subsection{The $V^{5,2}/\BZ_k$ theory} \label{sec:V52}
In this subsection, we focus on field theories dual to ${\rm AdS}_4 \times V^{5,2}/\BZ_k$, where $V^{5,2}$ is a homogeneous Sasaki-Einstein seven-manifold known as a Stiefel manifold.   The latter can be described as the coset $V^{5,2} =\SO(5)/\SO(3)$, whose supergravity solution \cite{Fabbri:1999hw} possesses an $\SO(5) \times \U(1)_R$ isometry.  There are two known descriptions of such field theories; one proposed by Martelli and Sparks \cite{Martelli:2009ga} and the other proposed by Jafferis \cite{Jafferis:2009th}.  In the following, we refer to these theories as Model I and Model II, respectively.  Below we analyse the solutions to the BAEs in detail and show the equality between the topological free energy of two theories.

\subsubsection{Model I}
The description for Model I was first presented in \cite{Martelli:2009ga}.  The quiver diagram is depicted below.
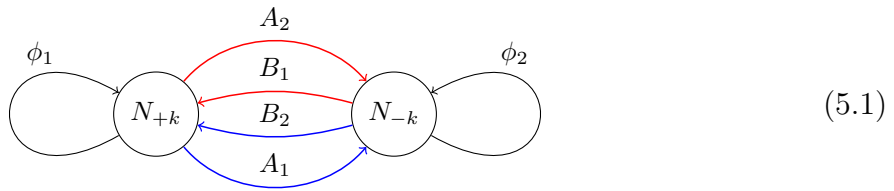
\bea
\begin{tikzpicture}[baseline, font=\footnotesize, scale=0.8]
\begin{scope}[auto,%
  every node/.style={draw, minimum size=0.5cm}, node distance=2cm];
  % the vertices
\node[circle] (USp2k) at (-0.1, 0) {$N_{+k}$};
\node[circle, right=of USp2k] (BN)  {$N_{-k}$};
\end{scope}
  % the edges
\draw[draw=blue,solid,line width=0.2mm,<-]  (USp2k) to[bend right=15] node[midway,above] {$B_2 $}node[midway,above] {}  (BN) ;
\draw[draw=blue,solid,line width=0.2mm,->]  (USp2k) to[bend right=50] node[midway,above] {$A_1$}node[midway,above] {}  (BN) ; 
\draw[draw=red,solid,line width=0.2mm,<-]  (USp2k) to[bend left=15] node[midway,above] {$B_1$} node[midway,above] {} (BN) ;  
\draw[draw=red,solid,line width=0.2mm,->]  (USp2k) to[bend left=50] node[midway,above] {$A_2$} node[midway,above] {} (BN) ;    
\draw[black,-> ] (USp2k) edge [out={-150},in={150},loop,looseness=10] (USp2k) node at (-2,1) {$\phi_1$} ;
\draw[black,-> ] (BN) edge [out={-30},in={30},loop,looseness=10] (BN) node at (5.8,1) {$\phi_2$};
\end{tikzpicture}
\eea
with the superpotential
\begin{equation}
 W = \Tr\left[ \phi_1^3 + \phi_2^3 +\phi_1(A_1 B_2 + A_2 B_1) + \phi_2 (B_2 A_1+ B_1 A_2) \right] \, .
\end{equation}

\paragraph{A solution to the BAEs.} Let us use the shorthand notation as in \eref{shorthandAB}. We look for a solution to BAEs, such that
\begin{align}\label{Delta constraint}
 \Delta_{\phi_i} + \Delta_{1} + \Delta_{4} =  2 \pi \, ,\qquad \qquad
 \Delta_{\phi_i} + \Delta_{2} + \Delta_{3} = 2 \pi \, , \qquad \qquad
 \Delta_{\phi_i} = \frac{2 \pi}{3} \, ,
\end{align}
and
\begin{align}\label{fluxconstraintV52}
 \fn_{\phi_i} + \fn_{1} + \fn_4 = 2\, ,\qquad \qquad
 \fn_{\phi_i} + \fn_{2} + \fn_3 = 2\, , \qquad \qquad
 \fn_{\phi_i} = \frac23 \, .
\end{align}

Observe that $\fn_{\phi_i}$ does not satisfy the quantisation condition $\fn_{\phi_i}\in \BZ$.  However, this problem can be cured easily by considering the twisted partition function on a Riemann surface $\Sigma_g$ of genus $g$ times $S^1$ \cite{Benini:2016hjo}. In this case, the flux constraints become
\begin{align}
 \fn_{\phi_i} + \fn_{1} + \fn_4 = 2(1-g)\, ,\qquad
 \fn_{\phi_i} + \fn_{2} + \fn_3 = 2(1-g)\, , \qquad
 \fn_{\phi_i} = \frac23 (1-g) \, .
\end{align}
By choosing $(1-g)$ to be an integer multiple of $3$, there always exists an integer solution to the above constraints.
As was pointed out in \cite{Benini:2016hjo}, the BAEs for the partition function on $\Sigma_g \times S_1$ (with $g>1$) is the same as that for $g=0$.
We can therefore solve the BAEs in the usual way.
 
The inner interval $[t_<, t_>]$ is given by
\be
t_< \text{ s.t. } \delta v(t_<) = - \Delta_3 \, ,\qquad\qquad t_> \text{ s.t. } \delta v(t_>) = \Delta_1 \, .
\ee
Outside the inner interval, we find that $\delta v(t) =  \tilde v(t) - v(t)$ is frozen to the constant boundary value $- \Delta_3$ ($ \Delta_1$) and it defines the left (right) tail.
Schematically:
\begin{center}
\begin{tikzpicture}[scale=2]
\draw (-1.5,0) -- (1.5,0);
\draw (-1.5,-.05) -- (-1.5, .05); \draw (-0.5,-.05) -- (-0.5, .05); \draw (0.5,-.05) -- (0.5, .05); \draw (1.5,-.05) -- (1.5, .05);
\node [below] at (-1.5,0) {$t_\ll$}; \node [below] at (-1.5,-.3) {$\rho=0$};
\node [below] at (-0.5,0) {$t_<$}; \node [below] at (-0.5,-.3) {$\delta v = -\Delta_3$};% \node [below] at (-0.5,-.6) {$Y_3 = 0$};
\node [below] at (0.5,0) {$t_>$}; \node [below] at (0.5,-.3) {$\delta v=\Delta_1$};% \node [below] at (0.5,-.6) {$Y_1 = 0$};
\node [below] at (1.5,0) {$t_\gg$}; \node [below] at (1.5,-.3) {$\rho=0$};
\end{tikzpicture}
\end{center}
The solution is as follows. The transition points are at
\be
\label{solution sum 2pi -- init -- V52}
t_\ll = - \frac{\mu}{k \Delta_3} \,,\qquad\quad t_< = - \frac{\mu}{k \Delta_4} \,,\qquad\quad t_> = \frac{\mu }{k \left( \frac{4 \pi}{3} - \Delta_3 \right)} \,,\qquad\quad t_\gg = \frac{\mu }{k \left( \frac{4 \pi}{3} - \Delta_4 \right)} \,.
\ee
In the left tail we have
\be
\begin{aligned}
\rho &= \frac{\mu +k \Delta_3 t}{\frac{2 \pi}{3} \left(\Delta_3-\Delta_4+\frac{4 \pi }{3}\right) \left(\Delta_4-\Delta_3\right)} \\[.5em]
\delta v &= - \Delta_3 \,,\qquad\qquad Y_3 = \frac{- k t\Delta_4 -\mu }{\Delta_4 - \Delta_3}
\end{aligned}
\qquad\qquad\qquad t_\ll < t < t_< \, .
\ee
In the inner interval we have
\be
\begin{aligned}
\rho &= \frac{2 \mu + k \left(\Delta_3+\Delta_4-\frac{4 \pi}{3}\right) t}{\frac{2 \pi }{3} \left[\left(\frac{4 \pi }{3}\right)^2-\left(\Delta_4 - \Delta_3\right)^2 \right]} \\[.5em]
\delta v &= -\frac{\left(\Delta_3+\Delta_4-\frac{4 \pi }{3}\right) \mu -  \frac{4 \pi}{3}k  \left(\Delta_3+\Delta_4\right) t+\left(\Delta_3^2+\Delta_4^2\right) t}{2 \mu +k \left(\Delta_3+\Delta_4-\frac{4 \pi }{3}\right) t}
\end{aligned}
\qquad\qquad t_< < t < t_>
\ee
and $\delta v'>0$.
In the right tail we have
\be
\begin{aligned}
\rho &= \frac{\mu - k \Delta_1 t}{\frac{2 \pi}{3} \left(\Delta_3-\Delta_4+\frac{4 \pi }{3}\right) \left(\Delta_4-\Delta_3\right)} \\[.5em]
\delta v &= \Delta_1 \,,\qquad\qquad Y_1 = \frac{- k t \left( \Delta_3 - \frac{4 \pi}{3}\right) - \mu}{\Delta_4 - \Delta_3}
\end{aligned}
\qquad\qquad\qquad t_> < t < t_\gg \, .
\ee
Finally, the normalization fixes
\be
\label{solution sum 2pi -- end -- V52}
\mu = \sqrt{k \left(\frac{4 \pi}{3}-\Delta_3\right) \Delta_3 \left(\frac{4 \pi }{3}-\Delta_4\right) \Delta_4} \, ,
\ee
with
\begin{equation}\label{Delta inequality -- V52}
 0 < \Delta_{3,4} < \frac{4 \pi}{3} \, .
\end{equation}
The solution satisfies
\begin{equation}
 \int dt\, \rho(t) \, \delta v(t) = 0 \, .
\end{equation}

We should take the solution to the BAEs and plug it back into the index. For higher genus $g$, formula \eref{genentropy} receives a simple modification, as discussed in \cite{Benini:2016hjo}, as follows,
\begin{align} \label{genentropyhigherg}
& \frac{\mathfrak{F}}{N^{3/2}} = - \frac{|G| \pi^2}{3}(1-g) \int dt\, \rho(t)^2 - \sum_{a=1}^{|G|} \ft_a \int dt\, t\, \rho(t) \nn \\
& + \frac{1}{2} \int dt\, |t|\, \rho(t) \left[\sum_{\substack{\text{anti-funds} \\ a}} (\tilde\fn_a - 1+g) + \sum_{\substack{\text{funds} \\ a}} (\fn_a - 1+g)\right] \nn \\
& - \int dt\, \rho(t)^2 \sum_{\substack{\text{bi-funds} \\ (b,a) \text{ and } (a,b) }} \left[(\fn_{(b,a)}-1+g)\, g'_+ \left(\delta v(t) + \Delta_{(b,a)}\right) + (\fn_{(a,b)}-1+g)\, g'_- \left(\delta v(t) - \Delta_{(a,b)}\right)\right] \nn \\
& - \sum_{\substack{\text{bi-fund} \\ (b,a) }} \fn_{(b,a)} \int_{\delta v \approx - \Delta_{(b,a)} (\text{mod } 2\pi)}  dt \, \rho(t) Y_{(b,a)}
- \sum_{\substack{\text{bi-fund} \\ (a,b) }} \fn_{(a,b)} \int_{\delta v \approx  \Delta_{(a,b)} (\text{mod } 2\pi)}  dt \, \rho(t) Y_{(a,b)} \, .
\end{align}

Doing the integration, we obtain the  following expression for the topological free energy,
\bea \label{entropy bi-fundamental}
\mathfrak{F} & = -\frac{2}{3}(1-g) \frac{k^{1/2} N^{3/2}}{\sqrt{\left(\frac{4 \pi}{3} - \Delta_3\right) \Delta_3 \left(\frac{4 \pi}{3} - \Delta_4\right) \Delta_4}}
 \Bigg\{\left(\frac{4 \pi}{3}-\Delta_3\right) \Delta_3 \left(\frac{2 \pi}{3} - \Delta_4\right) \frac{\fn_4}{1-g} \nn \\
 & + \Delta_4 \left[\left(\frac{2 \pi}{3}-\Delta_3\right) \left(\frac{4 \pi }{3} - \Delta_4\right) \frac{\fn_3}{1-g} - \frac{2 \Delta_3}{3} \left(\Delta_3+\Delta_4-\frac{8 \pi}{3}\right)\right]\Bigg\} \, .
\eea
We check that the topological free energy indeed satisfies the index theorem for this model on $\Sigma_g \times S^1$:
\begin{equation} \label{indexhigherg}
 \mathfrak{F} = (1-g) \left\{ - \frac{2}{\pi} \, \wb{\mathcal{V}}(\Delta_I) \,
 - \sum_{I}\, \left[ \left(\frac{\fn_I}{1-g} - \frac{\Delta_I}{\pi}\right) \frac{\partial \wb{\mathcal{V}}(\Delta_I)}{\partial \Delta_I}
  \right] \right\} \,,
\end{equation}
with
\begin{equation}
 \wb{\mathcal{V}}(\Delta_I) = \frac{2}{3} \mu N^{3/2} \, .
\end{equation}

\subsubsection{Model II}
The description for Model II was first presented in \cite{Jafferis:2009th}. The quiver diagram is depicted below.
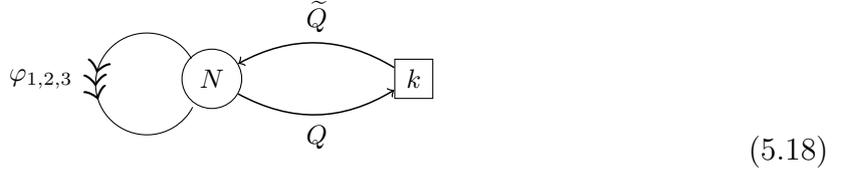
\bea
\begin{tikzpicture}[font=\footnotesize, scale=0.9]
\begin{scope}[auto,%
  every node/.style={draw, minimum size=0.5cm}, node distance=2cm];
\node[circle]  (UN)  at (0.3,1.7) {$N$};
\node[rectangle, right=of UN] (Ur) {$k$};
\end{scope}
\draw[decoration={markings, mark=at position 0.45 with {\arrow[scale=2.5]{>}}, mark=at position 0.5 with {\arrow[scale=2.5]{>}}, mark=at position 0.55 with {\arrow[scale=2.5]{>}}}, postaction={decorate}, shorten >=0.7pt] (-0,2) arc (30:335:0.75cm);
\draw[draw=black,solid,line width=0.2mm,->]  (UN) to[bend right=30] node[midway,below] {$Q$}node[midway,above] {}  (Ur) ; 
\draw[draw=black,solid,line width=0.2mm,<-]  (UN) to[bend left=30] node[midway,above] {$\tQ$} node[midway,above] {} (Ur) ;    
\node at (-2.2,1.7) {$\varphi_{1,2,3}$};
\end{tikzpicture}
\eea
We start from the superpotential
\begin{equation}\label{superpotential fundamental}
 W = \Tr \left\{ \varphi_3 \left[ \varphi_1, \varphi_2 \right] + \sum_{j=1}^{k} q_j \left( \varphi_1^2 + \varphi_2^2 + \varphi_3^2 \right) \tilde q^j \right\} \, .
\end{equation}
The $\SO(5)$ symmetry of $V^{5,2}$ can be made manifest by using the following variables \cite{Cremonesi:2016nbo}:
\bea
X_1 =  \frac{1}{\sqrt{2}}(\varphi_1 + i \varphi_2)\, , \qquad X_2 = \frac{1}{\sqrt{2}}(\varphi_1 - i \varphi_2)\, , \qquad X_3 = i \varphi_3 \, .
\eea
In terms of these new variables, the superpotential can be rewritten as
\bea
W = \Tr \left\{ X_3  [X_1 , X_2] +   \sum_{j=1}^k q_j ( X_1 X_2+ X_2 X_1 - X_3^2) \tilde q^j \right\}\, .
\eea

\paragraph{A solution to the BAEs.} The superpotential enforces
\begin{equation}
 \label{Delta constraint complex}
 \Delta_{X_1} + \Delta_{X_2} = \frac{4 \pi}{3}\, ,\qquad \qquad
 \Delta_{q_j} + \tilde \Delta_{q_j} = \frac{2 \pi}{3}\, ,\qquad \qquad
 \Delta_{X_3} = \frac{2 \pi}{3} \, ,
\end{equation}
and
\begin{equation}\label{flux constraint complex}
 \fn_{X_1} + \fn_{X_2} = \frac{4}{3}\, ,\qquad \qquad
 \fn_{q_j} + \tilde \fn_{q_j} = \frac{2}{3}\, ,\qquad \qquad
 \fn_{X_3} = \frac{2}{3} \, .
\end{equation}

As in the previous subsection, the quantisation conditions $\fn_I \in \BZ$ can be satisfied by considering the twisted partition function on $\Sigma_g \times S^1$.  The flux constraints are modified to be
\begin{equation}\label{flux constraint complex}
 \fn_{X_1} + \fn_{X_2} = \frac{4}{3}(1-g) \, ,\qquad
 \fn_{q_j} + \tilde \fn_{q_j} = \frac{2}{3}(1-g) \, ,\qquad 
 \fn_{X_3} = \frac{2}{3}(1-g) \, .
\end{equation}
Here we choose $(1-g)$ to be an integer multiple of $3$.  The solution to the BAEs are given below.

Setting to zero the variations with respect to $\rho(t)$, we find that the density is given by
\begin{equation}\label{density tm tp complex}
 \rho(t) = \frac{ \mu -\frac{2 \pi  k}{3} \left| t\right| + t \Delta_m}{\frac{2 \pi}{3} \left(\frac{4 \pi}{3} - \Delta_{X_1}\right) \Delta_{X_1}} \, .
\end{equation}
The support $[t_- , t_+]$ of $\rho(t)$ is determined by $\rho(t_\pm)=0$.  We obtain
\bea
 t_{\pm} =  \pm \frac{\mu }{\frac{2 \pi k}{3} \pm \Delta_m}\, .
\eea
Requiring that $\int_{t_{-}}^{t_{+}} \, dt \, \rho(t) =1$, we have
\bea
 \mu = \sqrt{\frac{\left(\frac{4 \pi}{3} - \Delta_{X_1}\right) \Delta_{X_1} \left[\left(\frac{2 \pi k}{3}\right)^2 - \Delta _m^2\right]}{k}} \, .
\eea

The topological free energy may then be found using \eref{genentropyhigherg}.  We obtain
\bea \label{entropy complex} 
\mathfrak{F} & =  \frac{2}{3} \frac{(1-g) N^{3/2}}{\sqrt{k \left(\frac{4 \pi}{3}-\Delta_{X_1}\right) \Delta_{X_1} \left[\left(\frac{2 \pi k}{3}\right)^2-\Delta_m^2\right]}} \times  \nn \\
& \qquad \Bigg\{\Delta_{X_1} \left[- \frac{\ft}{1-g} \Delta_m \left(\frac{4 \pi}{3}-\Delta_{X_1}\right) + \left( \frac{2 \pi k}{3} \right)^{2} \left(\frac{\Delta_{X_1}}{\pi} - 2\right)+\frac{2 \Delta_m^2}{3}\right]  \nn \\
& \qquad - \left(\frac{2 \pi}{3}-\Delta_{X_1} \right) \frac{{\fn}_{X_1}}{1-g} \left[\left(\frac{2 \pi k}{3}\right)^2-\Delta_m^2\right] \Bigg\} \, .
\eea
It can also be checked that this topological free energy satisfies \eref{indexhigherg}.

\paragraph{Matching with Model I.} By taking
\begin{equation} \label{matchDeltaIandII}
\Delta_{X_1} = \Delta_3 \, ,  \qquad \Delta_m = k \left(\frac{2 \pi }{3} -\Delta _4 \right) \, ,  \qquad
\fn_{X_1} = \fn_3 \, , \qquad  \ft = k \left[\frac{2}{3}(1-g) -\fn _4 \right] \, ,
\end{equation}
we see that Eq.\,\eqref{entropy complex} reduces to Eq.\,\eqref{entropy bi-fundamental}.

\subsection{The flavored ABJM theory}
Let us consider the flavored ABJM models studied in \cite{Benini:2009qs,Cremonesi:2010ae}
\bea
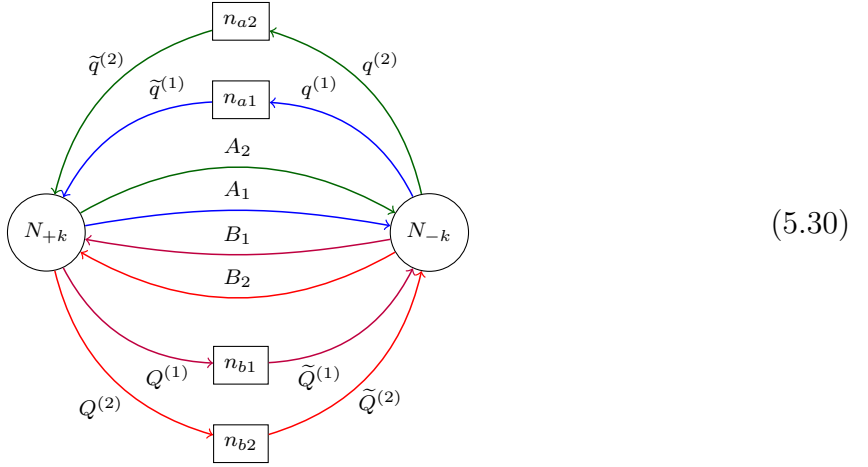
\begin{tikzpicture}[baseline, font=\scriptsize, scale=0.8]
\begin{scope}[auto,%
  every node/.style={draw, minimum size=0.5cm}, node distance=4cm];
  % the vertices
\node[circle] (UN1) at (0, 0) {$N_{+k}$};
\node[circle, right=of UN1] (UN2)  {$N_{-k}$};
\node[rectangle] at (3.2,2.2) (UNa1)  {$n_{a1}$};
\node[rectangle] at (3.2,3.5) (UNa2)  {$n_{a2}$};
\node[rectangle] at (3.2,-2.2) (UNb1)  {$n_{b1}$};
\node[rectangle] at (3.2,-3.5) (UNb2)  {$n_{b2}$};
\end{scope}
  % the edges
\draw[draw=red,solid,line width=0.2mm,<-]  (UN1) to[bend right=30] node[midway,above] {$B_2 $}node[midway,above] {}  (UN2) ;
\draw[draw=blue,solid,line width=0.2mm,->]  (UN1) to[bend right=-10] node[midway,above] {$A_1$}node[midway,above] {}  (UN2) ; 
\draw[draw=purple,solid,line width=0.2mm,<-]  (UN1) to[bend left=-10] node[midway,above] {$B_1$} node[midway,above] {} (UN2) ;  
\draw[draw=black!60!green,solid,line width=0.2mm,->]  (UN1) to[bend left=30] node[midway,above] {$A_2$} node[midway,above] {} (UN2) ;   
\draw[draw=purple,solid,line width=0.2mm,->]  (UN1)  to[bend right=30] node[midway,right] {}   (UNb1);
\draw[draw=purple,solid,line width=0.2mm,->]  (UNb1) to[bend right=30] node[midway,left] {} (UN2) ; 
\draw[draw=red,solid,line width=0.2mm,->]  (UN1)  to[bend right=30]  node[midway,right] {} (UNb2);
\draw[draw=red,solid,line width=0.2mm,->]  (UNb2) to[bend right=30] node[midway,left] {} (UN2); 
\draw[draw=blue,solid,line width=0.2mm,->]  (UN2)  to[bend right=30] node[pos=0.9,right] {}   (UNa1);
\draw[draw=blue,solid,line width=0.2mm,->]  (UNa1) to[bend right=30] node[pos=0.1,left] {} (UN1) ; 
\draw[draw=black!60!green,solid,line width=0.2mm,->]  (UN2)  to[bend right=30]  node[pos=0.9,right] {} (UNa2);
\draw[draw=black!60!green,solid,line width=0.2mm,->]  (UNa2) to[bend right=30] node[pos=0.1,left] {}  (UN1); 
\node at (4.5,-2.4) {$\tQ^{(1)}$};
\node at (2.,-2.4) {$Q^{(1)}$};
\node at (5.5,-2.8) {$\tQ^{(2)}$};
\node at (0.9,-2.9) {$Q^{(2)}$};
\node at (4.5,2.4) {$q^{(1)}$};
\node at (2.,2.4) {$\tilde{q}^{(1)}$};
\node at (5.5,2.8) {$q^{(2)}$};
\node at (1.,2.8) {$\tilde{q}^{(2)}$};
\end{tikzpicture}
\eea
with the superpotential
\bea\label{supflvABJM}
W &= \Tr \left( A_1 B_1 A_2 B_2 - A_1 B_2 A_2 B_1  \right) + \nn \\
& \quad \Tr\left[\sum_{j=1}^{n_{a1}} q_{j}^{(1)} A_1 \tilde q_{j}^{(1)}
 +\sum_{j=1}^{n_{a2}} q_{j}^{(2)} A_2 \tilde q_{j}^{(2)}
 +\sum_{j=1}^{n_{b1}} Q_{j}^{(1)} B_1 \tilde Q_{j}^{(1)}
 +\sum_{j=1}^{n_{b2}} Q_{j}^{(2)} B_2 \tilde Q_{j}^{(2)}\right]\, .
\eea

We adopt the notation as in \eref{shorthandAB} and denote by
\bea
\Delta_{ai} = \Delta_{q^{(i)}}~, \qquad \tilde{\Delta}_{ai} = \Delta_{\tilde{q}^{(i)}}~, \qquad {\Delta}_{bi} = \Delta_{{Q}^{(i)}}~, \qquad \tilde{\Delta}_{bi} = \Delta_{\tilde{Q}^{(i)}}~,
\eea
and similarly for $\fn_{ai}$ and $\fn_{bi}$.
The marginality of the superpotential implies
\begin{align}
 &\Delta_1 + \Delta_{a1} + \tilde \Delta_{a1}= 2\pi \, ,\qquad\qquad
 \Delta_2 + \Delta_{a2} + \tilde \Delta_{a2}= 2\pi \, ,\nn\\&
 \Delta_3 + \Delta_{b1} + \tilde \Delta_{b1}= 2\pi \, ,\qquad\qquad
 \Delta_4 + \Delta_{b2} + \tilde \Delta_{b2}= 2\pi \, ,
\end{align}
and
\begin{align}\label{flux constraint}
 &\fn_1 + \fn_{a1} + \tilde \fn_{a1}=2\, ,\qquad\qquad
 \fn_2 + \fn_{a2} + \tilde \fn_{a2}=2\, ,\nn\\&
 \fn_2 + \fn_{b1} + \tilde \fn_{b1}=2\, ,\qquad\qquad
 \fn_4 + \fn_{b2} + \tilde \fn_{b2}=2\, .
\end{align}

\subsubsection{A solution to the system of BAEs}

The large $N$ expression for the Bethe potential, using \eref{genbethepot}, can be written as
\begin{align}\label{large N Bethe potential -- Q111}
\frac{\mathcal{V}}{i N^{3/2}}&
=\int dt\, \rho(t)^2\, \sum\nolimits^*_a\left[\pm g_{\pm} \left(\delta v(t) \pm \Delta_a\right)\right]
+\int dt\, t\, \rho(t)\, \left(\Delta_m^{(2)} - \Delta_m^{(1)} \right)\nn\\&
-\frac{1}{2}\int dt\, |t|\, \rho(t)\,\left[\sum\nolimits^*_f (\pm n_f)\delta v(t)- \sum_{i=1}^{2}\left(n_{ai}\Delta_{i} + n_{bi}\Delta_{i+2}\right)\right] \nn \\%%
& -\frac{i}{N^{1/2}}\int dt\, \rho(t)\, \sum\nolimits^*_a\left[\pm \Li_2 \left(e^{i\left(\delta v(t)\pm \Delta_a\right)}\right)\right]
-\mu \left(\int dt\, \rho(t)-1\right) \, ,
\end{align}
where we introduced the notations
\begin{equation}
 \sum\nolimits^*_f=\sum_{\substack{f=a1,a2:+\\ f=b1,b2:-}}\, ,\qquad \qquad \sum\nolimits^*_a=\sum_{\substack{a=3,4:+\\ a=1,2:-}}\, .
\end{equation}

\paragraph{The solution for $k=0$ and $n_{a1}=n_{a2}=n\, ,\; n_{b1}=n_{b2}=0$.}
As pointed out in \cite{Benini:2009qs}, this theory is dual to $\mathrm{AdS}_4 \times Q^{1,1,1}/\mathbb{Z}_n$.
The manifold $Q^{1,1,1}$ is defined by the coset
\be
\frac{\SU(2) \times \SU(2) \times \SU(2)}{\U(1) \times \U(1)} \, ,
\ee
and has the isometry
\be
\SU(2) \times \SU(2) \times \SU(2) \times \U(1) \, .
\ee
Using the symmetries of the quiver, we set for simplicity
\begin{align}
 \Delta_1=\Delta_2=\pi-\Delta_3=\pi-\Delta_4=\Delta\, .
\end{align}

Let $\Delta_m$ be the following linear combination of the topological chemical potentials of the two gauge groups:
\bea 
\Delta_m = \Delta_m^{(1)} - \Delta_m^{(2)}\, .
\eea
Solving the BAEs equations, we obtain the following general solution
\bea
 \rho(t) & = -\frac{n \pi \left| t\right| + 2 \Delta_m\, t - 2 \mu}{\pi^3} \, ,\nn \\
 \delta v(t) & = \Delta + \frac{\pi \left(\mu - \Delta_m\, t \right)}{n \pi \left| t\right| + 2 \Delta_m\, t - 2 \mu}\, ,
\eea
on the support $[t_- , t_+]$.  We determine $t_\pm$ from $\delta v (t_\pm)=-(\pi-\Delta)$,
\begin{equation}
 t_- = -\frac{\mu }{n \pi - \Delta_m}\, , \qquad \qquad t_+ = \frac{\mu }{n \pi + \Delta_m} \, .
\end{equation}
The normalization $\int_{t_-}^{t_+} dt\, \rho(t)=1$ fixes
\begin{equation}
 \mu = \frac{\pi}{\sqrt{n}} \frac{\left| n^2 \pi^2 - \Delta_m^2\right|}{\sqrt{3 n^2 \pi^2-\Delta_m^2}} \, .
\end{equation}
The solution satisfies,
\begin{equation}
 \int dt\, \rho(t)\, \delta v(t) = \Delta - \frac{2 n^2 \pi^3}{3 n^2 \pi^2 - \Delta_m^2} \, .
\end{equation}

\subsubsection{The index at large $N$}

The matrix model for the topological free energy functional in this case reads
\begin{align}
 \frac{\mathfrak{F}}{N^{3/2}} & = - \int dt\, \rho(t)^2 \bigg[ \frac{2\pi^2}3 + \sum\nolimits^*_{a} (\fn_a-1) g_\pm'\big( \delta v(t) \pm \Delta_a \big) \bigg] \nn \\
  & - \frac{1}{2} \sum_{i=1}^2 \left(n_{ai} \fn_{i} + n_{bi} \fn_{i+2}\right) \int dt\, |t|\, \rho(t) - \left( \ft + \tilde \ft \right) \int dt\, t\, \rho(t) \nn \\
  & - \sum_{a=1}^4 \fn_a \int_{\delta v \,\approx\, \varepsilon_a \Delta_a} \hspace{-2em} dt\, \rho(t) \, Y_a(t) \, ,
 \end{align} 
where we have used the behavior
\be
\delta v(t) = \varepsilon_a \left( \Delta_a - e^{- N^{1/2} Y_a(t)} \right)\, , \qquad\qquad \varepsilon_a = (1,1,-1,-1) \, ,
\ee
in the tails.
For the theory dual to AdS$_4 \times Q^{1,1,1}/\mathbb{Z}_n$ we find
\bea
\mathfrak{F} = -\frac23 \frac{N^{3/2}}{\sqrt{n} \left(3 \pi ^2 n^2 - \Delta_m^2\right)^{3/2}} \left[\pi \left(\ft + \tilde\ft \right) \left(\Delta_m^3-5 \pi^2 \Delta_m n^2\right) + \Delta_m^4 - 3 \pi ^2 n^2 \left(\Delta_m^2-2 \pi ^2 n^2\right)\right]\, .
\eea

\subsection{$\U(N)$ gauge theory with adjoints and fundamentals}

In this section, we consider the following flavored toric quiver gauge theory \cite{Benini:2009qs}
\bea
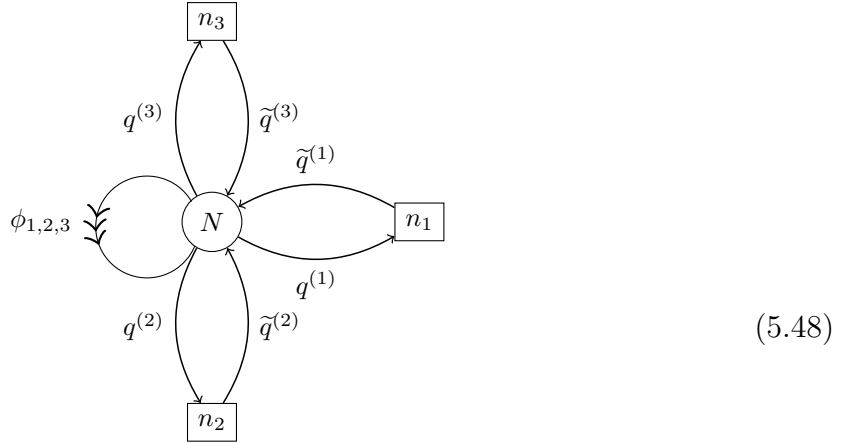
\begin{tikzpicture}[baseline, font=\footnotesize, scale=0.9]
\begin{scope}[auto,%
  every node/.style={draw, minimum size=0.5cm}, node distance=2cm];
\node[circle]  (UN)  at (0.3,1.7) {$N$};
\node[rectangle, right=of UN] (Ur1) {$n_1$};
\node[rectangle, below=of UN] (Ur2) {$n_2$};
\node[rectangle, above=of UN] (Ur3) {$n_3$};
\end{scope}
\draw[decoration={markings, mark=at position 0.45 with {\arrow[scale=2.5]{>}}, mark=at position 0.5 with {\arrow[scale=2.5]{>}}, mark=at position 0.55 with {\arrow[scale=2.5]{>}}}, postaction={decorate}, shorten >=0.7pt] (-0,2) arc (30:340:0.75cm);
\draw[draw=black,solid,line width=0.2mm,->]  (UN) to[bend right=30] node[midway,below] {$q^{(1)}$}node[midway,above] {}  (Ur1) ; 
\draw[draw=black,solid,line width=0.2mm,<-]  (UN) to[bend left=30] node[midway,above] {$\tilde{q}^{(1)}$} node[midway,above] {} (Ur1) ;    
\draw[draw=black,solid,line width=0.2mm,->]  (UN) to[bend right=30] node[midway,left] {$q^{(2)}$}node[midway,above] {}  (Ur2) ; 
\draw[draw=black,solid,line width=0.2mm,<-]  (UN) to[bend left=30] node[midway,right] {$\tilde{q}^{(2)}$} node[midway,above] {} (Ur2) ;  
\draw[draw=black,solid,line width=0.2mm,->]  (UN) to[bend left=30] node[midway,left] {$q^{(3)}$}node[midway,above] {}  (Ur3) ; 
\draw[draw=black,solid,line width=0.2mm,<-]  (UN) to[bend right=30] node[midway,right] {$\tilde{q}^{(3)}$} node[midway,above] {} (Ur3) ;    
\node at (-2.2,1.7) {$\phi_{1,2,3}$};
\end{tikzpicture}
\eea
with the superpotential
\begin{equation}\label{superpotentialU(n)}
 W = \Tr \left\{ \phi_1\left[\phi_2,\phi_3\right]
 + \sum_{j=1}^{n_1} q_j^{(1)} \phi_1 \tilde q_{j}^{(1)}
 + \sum_{j=1}^{n_2} q_j^{(2)} \phi_2 \tilde q_{j}^{(2)}
 + \sum_{j=1}^{n_3} q_j^{(3)} \phi_3 \tilde q_{j}^{(3)}\right\}\, .
\end{equation}

The marginality condition on the superpotential \eqref{superpotentialU(n)} implies that
\begin{align}\label{Delta constraintU(n)}
 \sum_{i=1}^{3} \Delta_{\phi_i} = 2 \pi\, ,  \qquad \Delta_{q_j^{(i)}} +\tilde \Delta_{q_j^{(i)}} + \Delta_{\phi_i} = 2 \pi \, ,
\end{align}
and
\begin{align}\label{flux constraintU(n)}
 &\sum_{i=1}^3 \fn_{\phi_i} = 2\, ,\qquad\qquad  \fn_{q_j^{(i)}} + \tilde \fn_{q_j^{(i)}} + \fn_{\phi_i} = 2 \, .
\end{align}
Let $\Delta_m$ and $\ft$ be the chemical potential and the background flux for the topological symmetry associated with the $\U(N)$ gauge group.

\paragraph{The solution.} On the support of $\rho(t)$, the solution is
\begin{equation}\label{density tm tp}
 \rho(t) = \frac{2 \left(\mu +t\, \Delta_m \right)-\left| t\right|  \bar{\Delta}}{2\hat\Delta } \, ,
\end{equation}
where we defined
\bea
 \hat\Delta = \prod_{f=1}^{3} \Delta_{\phi_f}\, , \qquad \bar{\Delta} = \sum_{f=1}^{3} n_f \Delta_{\phi_f}\, .
\eea
Let us denote by $[t_- , t_+]$ the support of $\rho(t)$.  We determine $t_\pm$ from the condition $\rho(t_\pm)=0$ and obtain
\bea
t_\pm = \pm \frac{2\mu}{\bar{\Delta} \mp 2 \Delta_m}\, .
\eea
The normalization $\int_{t_-}^{t_+} dt \, \rho(t) =1$ fixes the Lagrange multiplier $\mu$,
\begin{align}
 \mu & = \sqrt{ \frac{\hat\Delta}{2 \bar{\Delta}} \left( \bar{\Delta} - 2 \Delta_m\right) \left( \bar{\Delta} + 2 \Delta_m\right)} \, .
\end{align}
Using the same methods presented earlier, we obtain the following expression for the topological free energy,
\begin{align}\label{freeU(N)}
 \mathfrak{F} & = -\frac{N^{3/2}}{3}  \sqrt{ \frac{\hat\Delta}{2 \bar{\Delta}} \left( \bar{\Delta} - 2 \Delta_m\right) \left( \bar{\Delta} + 2 \Delta_m\right)} \Bigg[ \hat{\fn} +\frac{\bar{\fn} \left(\bar{\Delta}^2 + 4 \Delta_m^2\right)}{\bar{\Delta} \left(\bar{\Delta}^2-4 \Delta _m^2\right)} -\frac{8 \Delta_m}{\bar{\Delta}^2-4 \Delta _m^2}  \Bigg] \, ,
\end{align}
where
\bea
\hat{\fn} = \sum_{i=1}^3 \frac{\fn_{\phi_i}}{\Delta_{\phi_i}}\, , \qquad \bar{\fn} = \sum_{i=1}^3 n_i \fn_{\phi_i}\, .
\eea
When $n_1 = n_2 = 0$, and $n_3 = r$, the moduli space reduces to $\BC^2 \times \BC^2 / \BZ_r$ and
Eq.\,\eqref{freeU(N)} becomes the topological free energy of the ADHM quiver [see Eq.\,\eqref{freeADHM}].
This is consistent with the fact that this theory is dual to AdS$_4 \times S^7/\BZ_r$.

\section{Discussion and Conclusions} \label{sec:conclude}
In this paper, we study the topologically twisted index at large $N$ and fixed Chern-Simons levels for a number of three-dimensional $\cN \geq 2$ gauge theories with known M-theory duals.
Using the localization method, the index can be written as a contour integral of a meromorphic form, whose position of the poles is governed by a set of algebraic equations, dubbed as the Bethe ansatz equations (BAEs).
For each theory, we present explicitly the solution to the system of BAEs. The topological free energy, which is the real part of the logarithm of the twisted index, is then computed from such a solution.

In \cite{Hosseini:2016tor}, it has been shown that the Bethe potential for any $\cN \geq 2$ theory is exactly the free energy of the same theory on the three-sphere, up to a normalization.
Our findings for the Bethe potential of theories with $\cN = 2$ supersymmetry are indeed in agreement with the previously reported results for the $S^3$ free energy \cite{Herzog:2010hf,Jafferis:2011zi,Amariti:2011uw}.
Moreover, in the other cases our results give a prediction for the $S^3$ free energy that has not appeared before in the literature.
We would like to emphasize that for all the models considered in this paper, the topological free energy, which was obtained by evaluating the functional \eqref{genentropy} on the solution to the BAEs, is consistent with the robust index theorem \eqref{index theorem} which is derived in \cite{Hosseini:2016tor}.

Our solutions have a certain important feature that is worth pointing out here. For theories whose all Chern-Simons levels are zero,
the density of eigenvalue distribution is supported on one interval and the $\delta v$'s are frozen throughout that interval;
whereas for quiver gauge theories having nonzero Chern-Simons levels, the solution to the BAEs is separated into several intervals, each of which contributes nontrivially to the topological free energy.

For gauge theories with $\CN=4$ and $\CN=3$ supersymmetry, whose geometric moduli space is a symmetric product of two ALE singularities,
we find that their topological free energy can be written as that of the ABJM theory times a numerical factor,
which is equal to the square root of the ratio between the product of the orders of the singularities and the Chern-Simons coupling of the ABJM theory.

Along the way, we perform nontrivial checks of various dualities, including mirror symmetry between the ADHM quiver and the Kronheimer-Nakajima quiver,
$\SL(2,\BZ)$ duality between $\CN=3$ theory and the Kronheimer-Nakajima quiver, and duality between two models that are dual to M-theory on $\text{AdS}_4 \times V^{5,2}/\BZ_k$.

We also calculate the topological free energy for theories associated with homogeneous Sasaki-Einstein seven-manifolds $N^{0,1,0}$, $V^{5,2}$, and $Q^{1,1,1}$ which are appealing in the context of the AdS/CFT correspondence.  A natural future direction is to generalize the result of \cite{Benini:2015eyy}, where it was shown that the topological free energy of the ABJM theory reproduces the entropy of magnetically charged static BPS black holes in AdS$_4 \times S^7$. In particular, it would be of great interest to compare the topological free energy of theories in this paper with the entropy of supersymmetric asymptotically $\text{AdS}_4$ black holes in four-dimensional $\cN = 2$ gauged supergravity \cite{Halmagyi:2013sla,Halmagyi:2013qoa,Klemm:2016wng}.

\section*{Acknowledgements}
We are indebted to Stefano Cremonesi, Nick Halmagyi, and Anton Nedelin for a number of useful discussions. Special thanks go to Alberto Zaffaroni for commenting on the manuscript of this paper and for several illuminating conversations. We also thank the JHEP referee for emphasising the quantisation condition for the magnetic fluxes on $S^2$; this leads to the improvement of Section \ref{sec:V52} of version 2 of this paper.
SMH is supported in part by INFN.

\bibliographystyle{ytphys}
\bibliography{ref}

\end{document}